\documentclass[aps,amsfonts,pra,twocolumn,showpacs]{revtex4-1}
\usepackage{epsfig,amsmath,amssymb,bm,epsf,graphicx,psfrag}
\usepackage{color,subfigure}

\def\ket#1{\vert#1\rangle}
\def\ketbra#1{\vert#1\rangle\langle#1\vert}
\newcommand{\bc}{\begin{center}}
\newcommand{\ec}{\end{center}}
\newcommand{\be}{\begin{equation}}
\newcommand{\ee}{\end{equation}}
\newcommand{\bea}{\begin{eqnarray}}
\newcommand{\eea}{\end{eqnarray}}

\begin{document}

\title{Quantum spin systems for  measurement-based quantum computation}
\author{Tzu-Chieh Wei}
\affiliation{C. N. Yang Institute for Theoretical Physics and
Department of Physics and Astronomy, State University of New York at
Stony Brook, Stony Brook, NY 11794-3840, USA}

\date{\today}

\begin{abstract}
Measurement-based quantum computation is different from other approaches for quantum computation, in that everything needs to be done is only local measurement on a certain entangled state. It thus uses entanglement as the resource that drives computation. We give a pedagogical treatment on the basics, and then review some selected developments beyond graph states, including Affleck-Kennedy-Lieb-Tasaki states and more recent 2D symmetry-protected topological states.  
\end{abstract}

\maketitle

\section{Introduction: Measurement-Based Quantum Computation}
Quantum computation exploits massive parallelism of many qubits under the unitary evolution~\cite{NielsenChuang00}. Measurement is only necessary at the last step of reading out the result of computation. However, it was realized that local measurements alone on certain many-qubit entangled states, such as cluster states~\cite{Cluster}, can simulate unitary gates and hence universal quantum computation~\cite{Oneway}. The price to pay is the use of one spatial dimension as the simulated time direction and the entanglement as the enabling resource. For example, as will be explained below, an array of entangled qubits can be used to simulate a sequence of one-qubit gates. With some entanglement between with two such arrays, two-qubit gates like the Controlled-NOT (CNOT) gate can be simulated. 
 Measurement in quantum mechanics generally gives outcomes randomly, but quantum computation is to implement certain fixed gigantic unitary operation. How do we deal with such randomness to yield a ``deterministic'' unitary evolution? The framework of measurement-based quantum computation (MBQC)~\cite{Oneway2,ChildsLeungNielsen} is to exploit useful entanglement structure and to deal with randomness in measurement to enable quantum computation, and is the focus of this review. 
 
 The main MBQC framework was proposed by Rassendorf and Briegel in the ``One-way quantum computers'' using the cluster state~\cite{Oneway}. One of the important precursory works is the teleportation-based quantum computation by Gottesman and Chuang~\cite{GottesmanChuang}, which was used to provide an understanding of the cluster-state quantum computation using the valence-bond or projected entangled pair states (PEPS) by Verstraete and Cirac~\cite{Verstraete}. Subsequently, a framework using matrix-product states and PEPS called quantum computation in the correlation space was developed by Gross and Eisert~\cite{Gross}.
 
 Entanglement is the enabling resource in MBQC and thus it is important to know what aspect of entanglement makes computation possible. For example, it is natural to regard states with litte entanglement as being useless and those with high entanglement as potentially useful. Indeed, this is the case~\cite{VandenNest1,VandenNest2}. However,  even if a state
possesses a high amount of entanglement, the measurement outcome
is so random that no advantage over classical random guessing is gained~\cite{Gross1,Bremner}. This leads us to believe that {it is the structure of the entanglement that matters for quantum computation\/}.

With classification of entanglement into short-range and long-range in recent condensed-matter literatures~\cite{ChenGuWen2}, one may wonder whether that perspective can give us any insight for the resourcefulness of entanglement. At the moment, long-range entanglement is a property of intrinsic topological order~\cite{WenTO}, such as exhibited in the Kitaev's toric code~\cite{toric}, gapped spin liquids~\cite{Balents}, or fractional quantum Hall states~\cite{TsuiStormerGossard,Laughlin}. Their utility for quantum computation originates from a very different perspective, i.e. using anyonic properties to enable quantum gates~\cite{TQC}, and not much is focused on the entanglement structure, except from the point of view of  topological entanglement.  Most of the so-called resource states for MBQC belong to the so-called short-range entangled states, such as the cluster states and the Affleck-Kennedy-Lieb-Tasaki (AKLT) states.  However, we wish to point out that even local measurements can convert a short-range entangled state to a long-range entanglement state, as demonstrated in the work by Bravyi and Raussendorf, where a cluster state can be converted to a surface code (a planar version of the toric code)~\cite{BravyiRaussendorf}. This later led to the fault tolerance of MBQC using a 3d cluster states~\cite{RaussendorfAnn,RaussendorfNJP}, where each 2d slice is used to simulate a  time step in the operation of anyons  in a 2d surface code~\cite{RaussendorfHarrington,Fowler}. 

  Even though not much further progress has been made in connecting intrinsic topological order to MBQC~\cite{BravyiRaussendorf, Morimae}, recently a connection of the type of topological order protected by symmetry to MBQC was unveiled. This symmetry-protected topological (SPT) order possesses only short-range entanglement, as opposed to the intrinsic topological order mentioned above.
   It was recognized by Else et al.~\cite{ElseSchwarzBartlettDoherty} that both the 1d cluster and the 1d AKLT states, which are capable of supporting arbitrary single-qubit gates, belong to a 1D nontrivial SPT phase protected by $Z_2\times Z_2$ on-site symmetry. Moreover, any ground state in the entire phase also supports a protected identity gate operation that acts as a perfect wire for transmission of quantum information. But the ability for the entire SPT phases for arbitrary qubit or qudit gates is only established recently by Miller and Miyake~\cite{MillerMiyake} and Stephen et al.~\cite{ElseSchwarzBartlettDoherty}.
 However, the status of 2d at the moment has not been explored much, with some results on certain fixed-point wave functions for universal quantum computation~\cite{MillerMiyake16,Hendrik,MillerMiyake16b}, and certain result slightly beyond fixed points~\cite{WeiHuang}. Future development may strengthen the notion of computational phases in SPT and other phases of matter.

This is a very biased and selected review of several persectives of MBQC. There have been other reviews on MBQC and related subjects~\cite{Oneway2,RaussendorfWei,Kwek,Fujii}. The author tries to be as pedagogical as possible at least in the first half of the review, choosing subjects to build up to the second half, in which he also tries to give a review on recent developments relating SPT order and MBQC. 

We comment on why the framework of MBQC restricts to local measurements only.  If we are given the ability to perform any two-qubit measurement, then universal quantum computation can be achieved even with product states. Therefore MBQC separates entanglement as its resource and local measurements can only decrease the entanglement in the system. This is why the name ``one-way'' quantum computer was also used~\cite{Oneway}. However, one can also imagine a certain combination of local and limited two-qubit measurements to extend the framework, but we will not consider that in this review.

\section{One dimension}
\subsection{1d cluster states}
\label{sec:1dcluster}
We will start with the 1d cluster state~\cite{Cluster} and explain how local measurement can be used to simulate unitary gates. The explanation here follows the approach by Nielsen~\cite{Nielsen}.

\smallskip\noindent {\bf Two-qubit warm up}. 
Consider an arbitrary qubit state $|\psi\rangle_1=a|0\rangle_1 + b|1\rangle_1$, where the subscript $1$ is used to label the qubit, and another qubit in the state $|+\rangle_2=(|0\rangle_2+|1\rangle_2)/\sqrt{2}$ state. Note that $|0\rangle$ and $|1\rangle$ are the +1 and -1 eigenstates, respectively,  of the Pauli Z matrix $Z=\sigma^z$. In the following, the normalization, such as $1/\sqrt{2}$, may be dropped for ease of notation.
Imagine we have a Controlled-Z gate ${\rm CZ}_{mn}\equiv |0\rangle\langle0|_m\otimes \openone_n + |1\rangle\langle 1|_m\otimes Z_n$. Note that the ${\rm CZ}$ gate is actually symmetric, ${\rm CZ}_{mn}={\rm CZ}_{nm}$, as the nontrivial action on the two qubits is only a phase shift: $|11\rangle \rightarrow -|11\rangle$.  This can come from, e.g., an Ising interaction in the presence of an external field.

Applying the ${\rm CZ}$ gate to the two qubits: $\big(a|0\rangle_1 + b|1\rangle_1\big)|+\rangle_2 \rightarrow |\Psi\rangle_{12}=a|0+\rangle + b |1-\rangle$, wehre $|-\rangle\equiv (|0\rangle_2-|1\rangle_2)/\sqrt{2}$\,; an entanglement is thus created between qubits 1 and 2. Imagine we perform a projective measurement on the first qubit, described by the observable $\hat{O}(\xi)= \cos\xi \, X + \sin\xi\, Y$, where $X=\sigma^x$ and $Y=\sigma^y$ are the Pauli X and Y matrices, respectively. An equivalent description of the measurement is the eigenstates of the observable $|\pm\xi\rangle\equiv ( |0\rangle \pm e^{i\xi}|1\rangle)/\sqrt{2}$, with eigenvalues $\pm1=(-1)^s$ (or equivalently a binary variable $s=0,1$) to describe the measurement outcome.

\begin{figure}
    \includegraphics[width=0.5\textwidth]{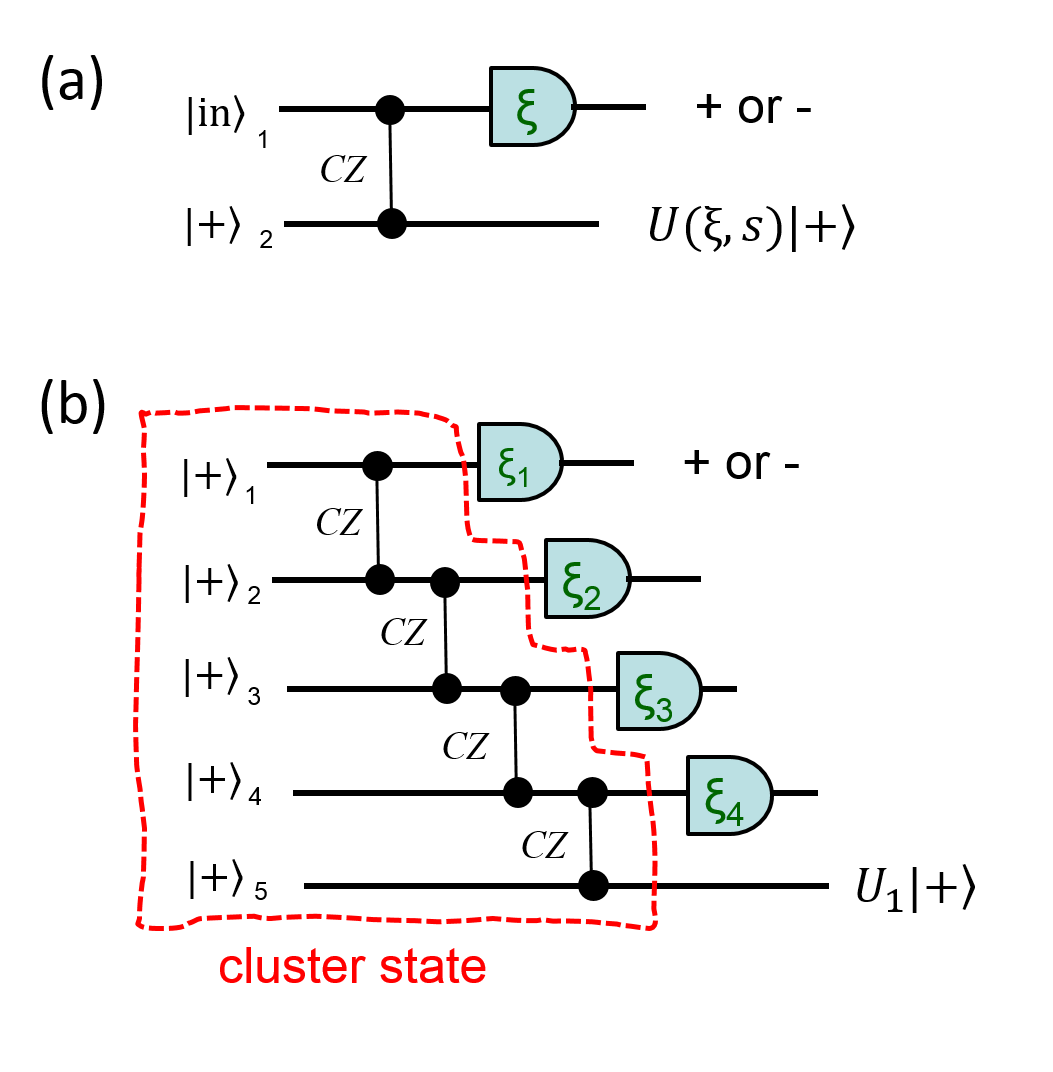}
  \caption{\label{fig:gateTeleportation}
  Gate teleportation: (a) a single one with an arbitrary input state $|\rm in\rangle$, (b) cascade of the procedure four times, but with $|+\rangle$ as the first input. The gate from the cascade is $U_1(\{\xi\})=\prod_{j=1}^4U(\xi_j,s_j)=U(\xi_4,s_4)U(\xi_3,s_3)U(\xi_2,s_2)U(\xi_1,s_1)$. The 1d cluster state with open boundary condition is indicated inside the red dashed box.}
\end{figure}

Depending on the measurement outcome $s$ on the first qubit, the second qubit is projected to 
\begin{equation}
|\psi'\rangle_2=\langle \pm \xi |\Psi\rangle_{12}\sim H\,e^{i\xi Z/2}\,Z^s (a|0\rangle_2+b|1\rangle_2),
\label{eqn:psi'}
\end{equation}
where an overall phase factor is omitted. Such a procedure of (1) entangling an arbitrary input qubit $|{\rm in}\rangle$ with a fixed $|+\rangle$, followed by (2) measuring the first qubit in $|\pm \xi\rangle$ basis, results in the quantum information $|{\rm in}\rangle$ teleported to the second qubit, with an additional outcome-dependent unitary gate $U(\xi,s)=H\,e^{i\xi Z/2}\,Z^s$, where $H$ is not a Hamiltonian but the so-called Hadamard gate
\begin{equation}
H=\frac{1}{\sqrt{2}}\begin{pmatrix}1 & 1\cr 1& -1\end{pmatrix}.
\end{equation}
This is referred to as the gate teleportation, schematically shown in Fig.~\ref{fig:gateTeleportation}.

\smallskip\noindent {\bf Arbitrary single-qubit gate implemented on a segment of qubits}. We also illustrate with a cascade of four such procedures and result in a gate $U_1(\{\xi\})=\prod_{j=1}^4U(\xi_j,s_j)=U(\xi_4,s_4)U(\xi_3,s_3)U(\xi_2,s_2)U(\xi_1,s_1)$. Suppose we would like to implement an SU(2) rotation described by the decomposition into Euler-angle rotations: $R(\alpha,\beta,\gamma)=e^{-i\alpha X/2}\,\,
e^{-i\beta Z/2}\, \,e^{-i\gamma X/2} $, we will see that this can be deterministically implemented up to some byproduct operators which are simple $X$ or $Z$. Take $\xi_1=0$ for simplicity. By using the fact that under the conjugation of the Hadamard gate, $X$ and $Z$ interchange, as well as that $X$ and $Z$ anticommute, we obtain that
\begin{eqnarray}
U_1(\{\xi\},\{s\})&=& Z^{s_1+s_3} X^{s_2+s_4} \,
e^{i (-1)^{s_1+s_3} \xi_4 X/2}\nonumber\\
&& e^{i (-1)^{s_2}\xi_3 Z/2}\, \,e^{i (-1)^{s_1}\xi_2 X/2}.
\label{eqn:U1}
\end{eqnarray} 
We see that the angles are not fixed but depend on outcomes $s_i$'s. However, if one realizes that one has the freedom to choose the subsequent measurement angles conditioned on the prior outcomes, we can achieve an almost deterministic rotation, $R(\alpha,\beta,\gamma)$, if we take
\begin{equation}
\xi_2=-(-1)^{s_1} \gamma, \ \xi_3=-(-1)^{s_2} \beta, \ \xi_4=-(-1)^{s_1+s_3}\alpha,
\end{equation}
and we obtain that
\begin{equation}
U_1(\{\xi\},\{s\})= Z^{s_1+s_3} X^{s_2+s_4} R(\alpha,\beta,\gamma).
\end{equation}

We see that there is a temporal ordering in the measurement, e.g. the  measurement of qubit 3 cannot take place before that of qubit 2, as its actual measurement axis $\xi_3$ depends on $s_2$, the outcome of measurement on qubit 2. We also have the trace of the randomness from the measurement, manifest in the byproduct operator $O_B\equiv Z^{s_1+s_3} X^{s_2+s_4}$.  But if this is last step before final readout, i.e., measurement in the fixed basis defined by observable $Z$, then $Z^{s_1+s_3}$ only affects the phase but not the classical reading of outcome $0$ or $1$, which may be flipped by $X^{s_2+s_4}$ but can be corrected by hand.
If this is not the end of gate operation, we can propage this to the next desirable unitary $V$, e.g., by implementing instead $O_B V O_B^{-1}$ to cancel the byproduct $O_B$ and then accumulating further byproduct operators. We repeat this until the last step when we need to read out the final outcome. The byproduct operator that can affect the readout is $X$, but we can correct this by flipping the outcome.

\smallskip \noindent {\bf Cluster states and parent Hamiltonians}. Since we can implement arbitrary rotation, we may as well let the input state to be $|+\rangle$ as well. Then we can separate all the CZ gates acting on all $|+\rangle$'s and define a 1d  cluster state (with an open-boundary condition)
\begin{equation}
|{\cal C}_{\rm open}\rangle=\left(\prod_{j=1}^{n-1} {\rm CZ}_{j,j+1}\right) |+\rangle_1|+\rangle_2\cdots |+\rangle_n.
\end{equation}
Such a cluster state allows us to simulate a sequence of arbitrary single-qubit rotations. We thus see that the one dimensionality serves as a discrete time direction of an effective qubit evolution.

One can also define a 1d cluster state on a periodic boundary condition,
\begin{equation}
|{\cal C}_{\rm peri}\rangle=\prod_{j=1}^{n} {\rm CZ}_{j,j+1} |+\rangle_1|+\rangle_2\cdots |+\rangle_n,
\end{equation}
where $n+1$ is identified as 1.

Since trivially that $|+\rangle$ is the +1 eigenstate of $X$, i.e., $X|+\rangle=|+\rangle$, we see that
\begin{equation}
\prod_{j=1}^{n} {\rm CZ}_{j,j+1} X_k = Z_{k-1} X_k Z_{k+1} \prod_{j=1}^{n} {\rm CZ}_{j,j+1},
\label{eqn:CZ-X}
\end{equation}
and hence ${\cal K}_k\equiv Z_{k-1} X_k Z_{k+1}$ is a special kind of operator, called the stabilizer operator, such that ${\cal K}_k |{\cal C}_{\rm peri}\rangle=|{\cal C}_{\rm peri}\rangle$. ${\cal K}$'s at different sites commute: $[{\cal K}_j,{\cal K}_i]=0$, so we can define a Hamiltonian 
\begin{equation}
\label{eqn:Hcluster}
\hat{H}_{\rm cluster}=-\sum_k {\cal K}_{k}=-\sum_k Z_{k-1} X_k Z_{k+1},
\end{equation}
such that $|{\cal C}_{\rm peri}\rangle$ is the unique ground state with a gap $\Delta=2$. For the open-boundary cluster state $|{\cal C}_{\rm open}\rangle$ its parent Hamiltonian differs from  Eq.~(\ref{eqn:Hcluster}) in the boundary terms: $-X_1 Z_2$ and $-Z_{n-1}X_n$.

It turns out that the cluster states $|{\cal C}_{\rm peri}\rangle$ and $|{\cal C}_{\rm open}\rangle$ are examples of the so-called symmetry-protected topological (SPT) states, protected by $Z_2\times Z_2$ symmetry (generated by $\dots I\otimes X \otimes I\otimes X\dots$ and $\dots X\otimes I \otimes X\otimes I\dots$). Another prominent example is the 1d spin-1 AKLT state, which we will discuss later.

\begin{figure}
   \includegraphics[width=0.4\textwidth]{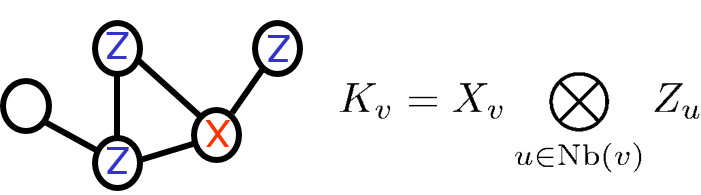}
  \caption{\label{fig:graphState}
  Illustration of a graph state and one of its stabilizer operators. ${\rm Nb}(u)$ indicates the set of neighboring vertices of vertex $u$. }
\end{figure}

We can in fact generalize the cluster state to the so-called graph state, illustrated in Fig.~\ref{fig:graphState}, where a qubit resides on each vertex and each edge of the graph indicates an CZ gate:
\begin{equation}
|{G}\rangle\equiv \left(\prod_{\langle i,j\rangle \in E} {\rm CZ}_{ij}\right) \mathop{\otimes}_{k\in V} |+\rangle_k,
\end{equation}
where $E$ indicates the set of edges and $V$ the set of vertices of the graph $G$.  Then there is a stabilizer operator for each vertex $u$,
\begin{equation}
\label{eqn:Ku}
{\cal K}_u =X_u \prod_{v, \langle u,v\rangle \in E} Z_v.
\end{equation} 
It is worth pointing out that under the unitary $U\equiv \prod_{\langle i,j\rangle \in E} {\rm CZ}_{ij}$, the Hamiltonian $\hat{H}_{\rm cluster}$ is transformed into a simple non-interacting one $H_{\rm para}=-\sum_k X_k$.

\begin{figure}
   \includegraphics[width=0.4\textwidth]{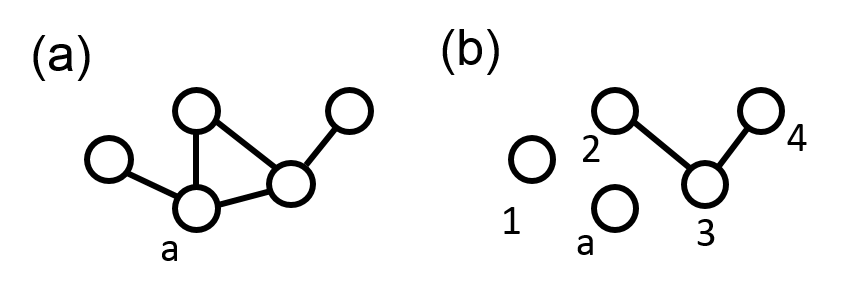}
  \caption{\label{fig:graphZ}
  Illustration of a graph state and measurement of qubit $a$ in the Z basis. }
\end{figure}

\smallskip \noindent {\bf Pauli Z measurement}. Let consider the effect of measuring a qubit, e.g. $a$, in the graph state in the Z basis. As illustrated in Fig.~\ref{fig:graphZ}, we will see that the effect is another graph state whose graph is obtained  by removing the vertex $a$ and all edges incident on it.
To prove this, we first note that we can imagine building up the graph state by applying the CZ gates to between all qubit pairs except those involving $a$, and we have a graph state $|\Psi_{G\backslash a}\rangle$ without the vertex $a$.  Thus carrying the last step of remaining CZ gates to obtain the graph state $|G\rangle$ is shown as follows,
\begin{equation}
|\Psi_{G}= |0\rangle_a |\Psi_{G\backslash a}\rangle + |1\rangle_a \left(\prod_{b \in {\rm Nb}(a)} Z_b\right) |\Psi_{G\backslash a}\rangle,
\end{equation}   
where ${\rm Nb}(a)$ denotes the set of $a$'s neighbors, i.e. $\{b|\langle a,b\rangle \in E\}$.
Thus measuring in the Z basis, we either obtain $|0\rangle_a$ and the remaining qubits are projected to $|\Psi_{G\backslash a}\rangle$ or $|1\rangle_a$ wth the remaining qubits being  $ \left(\prod_{b \in {\rm Nb}(a)} Z_b\right) |\Psi_{G\backslash a}\rangle$, which is a graph state, up to the product of Pauli Z operators.

\begin{figure}
   \includegraphics[width=0.3\textwidth]{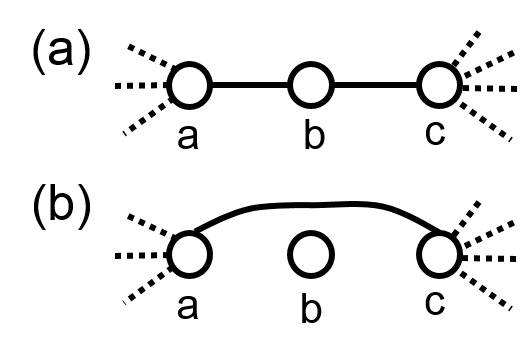}
  \caption{\label{fig:graphY}
  Illustration of a graph state and measurement of qubit $b$ in the Y basis. }
\end{figure}
\smallskip \noindent {\bf Pauli Y measurement}. Consider the graph in Fig.~\ref{fig:graphY}a and we want to perform a Y measurement on qubit $b$ of the graph state. The state of the remaining qubits is still a graph state with the graph depicted in Fig.~\ref{fig:graphY}b, up to local phase gates
\begin{equation}
S^{\pm 1}=\begin{pmatrix}
1 & 0 \cr
0 & \pm i
\end{pmatrix}.
\end{equation}
We will set up the steps for an elementary proof.  We will represent the graph state of the qubits connected to the left of qubit $a$ as $|\psi_a\rangle$, and the associated state with additional $Z$ gates acted as 
\[ |\tilde{\psi}_a\rangle= \prod_{b, \langle a,b\rangle \in E} Z_b |\psi_a\rangle
\] 
and similarly for those qubits connected to the right-hand side of qubit $c$.
The whole graph state is then 
\begin{eqnarray}
\!\!\!\!\!\!\!\!\!\!|G\rangle&=&|0\rangle_b (|0\rangle_a|\psi_a\rangle + |1\rangle_a|\tilde{\psi}_a\rangle) (|0\rangle_c|\psi_c\rangle + |1\rangle_c|\tilde{\psi}_c\rangle) \nonumber \\
\!\!\!\!\!\!\!\!\!\!&+&|1\rangle_b (|0\rangle_a|\psi_a\rangle - |1\rangle_a|\tilde{\psi}_a\rangle) (|0\rangle_c|\psi_c\rangle - |1\rangle_c|\tilde{\psi}_c\rangle).
\end{eqnarray}
Measuring in the Y basis corresponds to projecting qubit $b$ by $\langle \pm i|\equiv\langle 0| \mp i \langle 1|$ and it yields,
\begin{eqnarray}
|G'\rangle&=&{}_b\langle \pm i|G\rangle \\
&=& (|0\rangle_a|\psi_a\rangle + |1\rangle_a|\tilde{\psi}_a\rangle) (|0\rangle_c|\psi_c\rangle + |1\rangle_c|\tilde{\psi}_c\rangle) \nonumber \\
\!\!\!\!\!\!\!\!\!\!&&\mp i (|0\rangle_a|\psi_a\rangle - |1\rangle_a|\tilde{\psi}_a\rangle) (|0\rangle_c|\psi_c\rangle - |1\rangle_c|\tilde{\psi}_c\rangle).\nonumber\\
\!\!\!\!\!\!\!\!\!\! &=& (1\mp i)S^{\pm1}_a\otimes S^{\pm 1}_c \,{\rm CZ}_{ac} |G\backslash b\rangle, \label{eqn:Gb}
\end{eqnarray}
where $|G\backslash b\rangle$ is the graph state without qubit $b$.
It is seen that the effect of measuring in the Z and Y (as well as X) basis on a graph state can be represented by modifying the original graph; for more detailed discussions, see Refs.~\cite{Hein,Elliott}, including discussions on X measurement and elegant proof using the stabilizer formalism. For Z measurement, it is just deleting the vertex and incident edges. For Y measurement, the operation on the graph is the so-called local complementation on the measured vertex, followed by removing it and incident edges. But what we have shown above are sufficient for our applications below.

\smallskip\noindent {\bf Dynamics from the Hamiltonian}. We remark that most of the discussions in MBQC assume the measurements are done on a quantum state which has no dynamics, i.e., there is no Hamiltonian. In the case of the cluster state, we can still perform the simulations of single-qubit gates in the presence of the Hamiltnoian $\hat{H}_{\rm cluster}$, provided the measurements are done very quickly but only at  time intervals $2\pi m$, as $e^{-i 2\pi\,m \hat{H}_{\rm cluster}}=\openone$. The measurement will produce excitations that will evolve under  $e^{-i t \hat{H}_{\rm cluster}}$ but recohere at $t=2\pi m$.

\smallskip \noindent {\bf Valence-bond picture\/}.  Verstraete and Cirac provided an alternative picture of the 2d cluster-state quantum computation~\cite{Verstraete}. But here we illustrate it using the 1d cluster state. Such as valence-bond picture of a quantum state is also called a projected-entangled-pair state (PEPS)~\cite{PEPS}, and in the 1d case is the so-called matrix-product state (MPS)~\cite{MPS1,MPS2}. This perspective is based on the gate teleportation approach by Gottesman and Chuang~\cite{GottesmanChuang}.

In such as valence-bond picture, each site consists of two virtual qubits. One of the virtual qubit forms with one virtual qubit in the neighboring site in a valence bond of the form ${\rm CZ}|+\rangle|+\rangle=|0\rangle|+\rangle + |1\rangle|-\rangle$, where we have suppressed the normalization. But to make each site to a physical qubit, one takes the projection: \[
P_v=|0\rangle \langle 00| +|1\rangle  \langle 11|.
\]

\begin{figure}
   \includegraphics[width=0.48\textwidth]{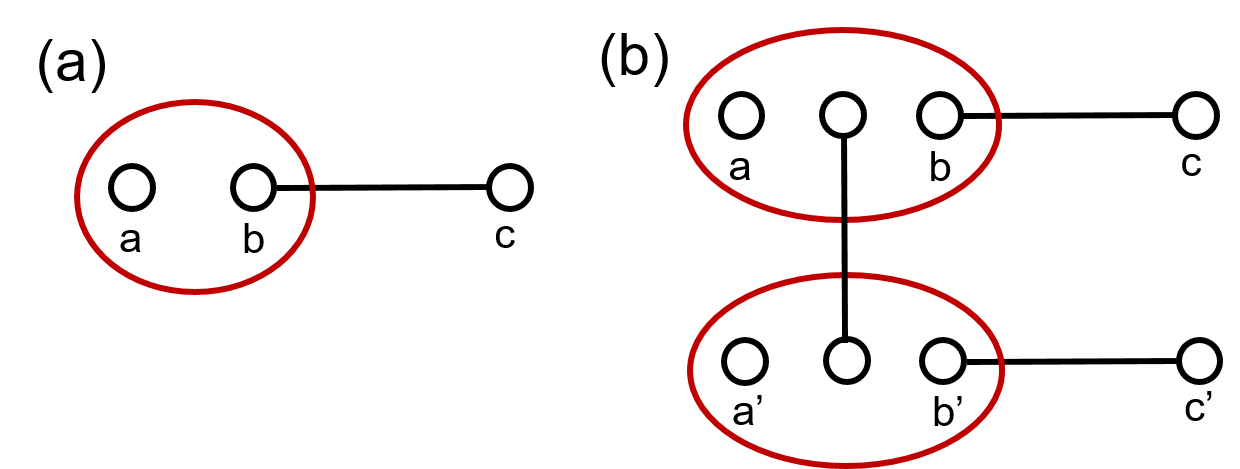}
  \caption{\label{fig:vbTeleportation}
 Valence-bond picture and teleportation.  (a) One qubit gate $|\psi_a\rangle\rightarrow U_{\rm 1-qubit}|\psi_c\rangle$ by measuring the joint qubits $a$ and $b$; the quantum state residing on qubit $a$ is teleported to $c$ and a single-qubit gate is applied. (b) Two qubit gate on $|\psi_{aa'}\rangle$ is achieved by measuring separately the system indicated by the two red circles and the resulting quantum state on $c$ and $c'$ is $|\psi_{cc'}\rangle=U_{\rm 2-qubit}|\psi_{aa'}\rangle$.}
\end{figure}

Imagine the virtual qubit $a$ encodes a state $|\psi\rangle_a=(\alpha|0\rangle_a+\beta|1\rangle_a)$; see Fig.~\ref{fig:vbTeleportation}a. Qubits $b$ and $c$ share the entanglement $|\psi\rangle_{bc}=|0\rangle|+\rangle + |1\rangle|-\rangle$.
A measurement on the physical qubit in the $|\pm \xi\rangle$ basis corresponds to a projection on the two virtual qubits,
\begin{equation}
\label{eqn:xiPv}
\langle \pm \xi| P_v =  \langle 00|_{ab} \pm \langle 11|_{ab} e^{- i \xi}.
\end{equation} 
This corresponds to a Bell measurement on qubits $a$ and $b$ but with only two outcomes $|\pm \xi\rangle$. After measurement, the state of $|\psi\rangle_a$ will be teleported to qubit $c$ up to a local unitary $U(\xi,s)=H\,e^{i\xi Z/2}\,Z^s$, which turns out to be  the same gate implemented in Eq.~(\ref{eqn:psi'}). By extending this to more qubits involved, a two-qubit gate can be realized by teleportation, as illustrated in Fig.~\ref{fig:vbTeleportation}b.

\begin{figure}
   \includegraphics[width=0.45\textwidth]{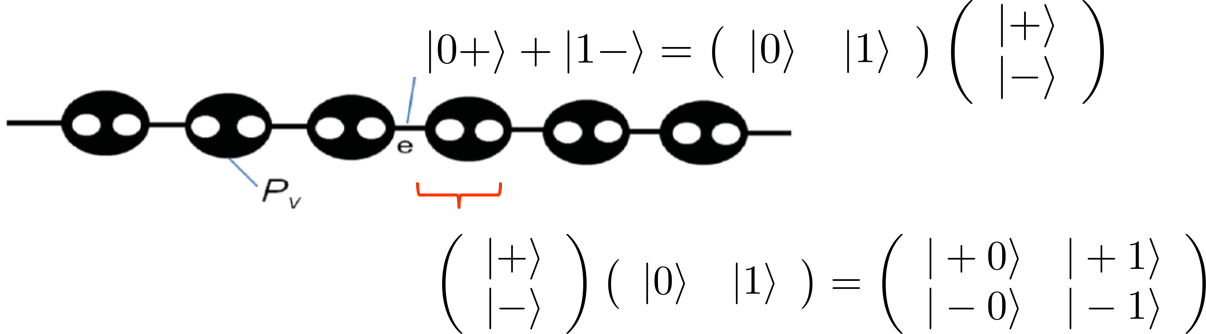}
   \\
   \includegraphics[width=0.45\textwidth]{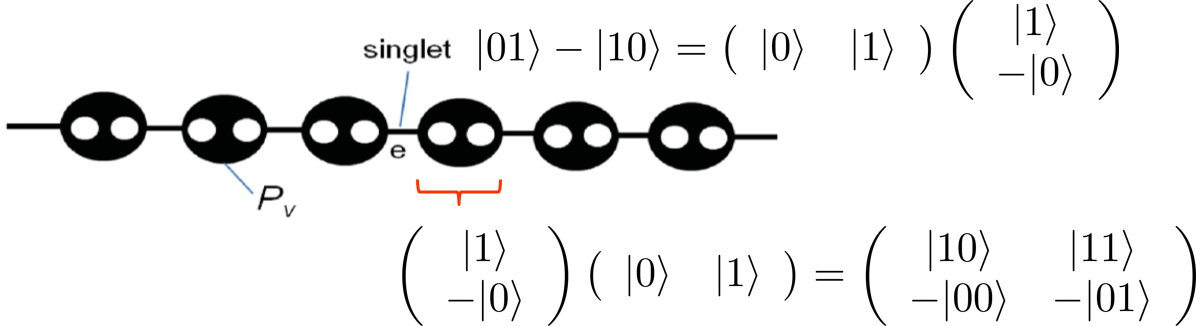}
  \caption{\label{fig:vbsCluster}
  Valence-bond picture of the cluster state (top) and the AKLT state (bottom). }
\end{figure}

\smallskip \noindent {\bf Matrix-product-state picture}. Refering to Fig.~\ref{fig:vbsCluster} we derive the so-called MPS representation. The first step is the local description in terms of the virtual qubits:
\[
\left(\begin{array}{c}
|+\rangle \\
|-\rangle
\end{array}\right)
\left(\begin{array}{cc}
|0\rangle & |1\rangle \end{array} \right) =
\left(\begin{array}{cc}
|+0\rangle & |+1\rangle \\
|-0\rangle & |-1\rangle\end{array} \right).
\]
This is followed by a projection $P_v$:

\begin{eqnarray}
&& P_v
\left(\begin{array}{cc}
|+0\rangle & |+1\rangle \\
|-0\rangle & |-1\rangle\end{array} \right)=\frac{1}{\sqrt{2}}\left(\begin{array}{cc}
|0\rangle & |1\rangle \\
|0\rangle & -|1\rangle\end{array} \right)\\
&&=|0\rangle A[0] + |1\rangle A[1]\\
&&=|0\rangle \left(|+\rangle\langle 0|\right)+|1\rangle \left(|-\rangle\langle 1|\right),
\end{eqnarray}
where we have abused the bracket notation for the two matrices: $A[0]=|+\rangle\langle 0|$ and $A[1]=|-\rangle\langle 1|$. The whole cluster state is then
\begin{equation}
|{\cal C}_{\rm peri}\rangle=\sum_{s_1,...,s_n}{\rm Tr}(A[s_1]A[s_2]\dots A[s_n])|s_1,\dots,s_n\rangle, 
\end{equation}
expressed in the MPS representation, where we have taken the periodic boundary condition. In general, we can use an open-boundary description by using two boundary vectors, e.g. $|R\rangle$ and $|L\rangle$, so that
\begin{equation}
|{\cal C}\rangle=\sum_{s_1,...,s_n}\big(\langle L|A[s_1]A[s_2]\dots A[s_n]|R\rangle\big)|s_1,\dots,s_n\rangle.
\end{equation}
When $|R\rangle= (1,1)^T$ and $|L\rangle=(1,0)^T$, then $|{\cal C}\rangle=|{\cal C}_{\rm open}\rangle$.
So local measurement on $n$-th spin with outcome described by a projector $|\phi\rangle\langle \phi|$, yields a transformation $|R\rangle \rightarrow A[\phi] |R\rangle$, where $A[\phi]=\langle \phi|0\rangle A[0]+\langle\phi|1\rangle A[1]$. More gates will be applied as spins on $(n-1)$-th, $(n-2)$-th, etc. are measured subsequently. This is the essence of the quantum computation in the correlation space by Gross and Eisert~\cite{Gross}. 
As before if we measure in the basis $|\pm \xi\rangle$, the resulting matrix is
\begin{equation}
A(\xi,s)=|+\rangle\langle 0|+(-1)^s e^{-i\xi}|-\rangle\langle 1|
=e^{-i\xi/2} H e^{i \xi Z/2} Z^{s},
\end{equation}
which is the same unitary gate $U(\xi,s)$ as in the gate teleportation case.
As we measure a sequence of sites, the product of $U$'s can be used to contruct arbitrary one-qubit gates. It is interesting to see that the quantum computation using the cluster state has the identical form in both the physical Hilbert space and the correlation space. 
\subsection{1d AKLT state}
The 1d AKLT state~\cite{AKLT,AKLT2} was invented to support Haldane's hypothesis on the spectral gap of integer spin chains with rotational symmetry~\cite{Haldane,Haldane2}. It is a prominent example of a 1d nontrivial SPT state. But it was also recognized that it could be used to simulate arbitrary single-qubit gates~\cite{Gross,Brennen}.  
We can understand why it is useful for simulating gates from two perspectives: (1) MPS picture, and (2) the quantum state reduction~\cite{ChenZeng}. The latter picture will be useful when we generalize to 2d AKLT states.

The bond state used to form the AKLT state is the singlet $|01\rangle-|10\rangle$ and the projection from virtual qubits to the physical spin is $P_v=|+1\rangle \langle 00| +|0\rangle (\langle 01|+\langle10|)/\sqrt{2} +|-1\rangle \langle 11|$. Similar to the derivation in the cluster state, we have
\begin{eqnarray}
&&P_v
\left(\begin{array}{cc}
|10\rangle & |11\rangle \\
-|00\rangle & -|01\rangle\end{array} \right)=\left(\begin{array}{cc}
|0\rangle/\sqrt{2} & |-1\rangle \\
-|+1\rangle & -|0\rangle\sqrt{2}\end{array} \right)\nonumber\\
&=&\frac{1}{\sqrt{2}}(|x\rangle X + |y\rangle Y + |z\rangle Z),
\label{eqn:PvAKLT}
\end{eqnarray}
where a convenient basis other than $|S_z=\pm1,0\rangle$ is 
defined by $|x\rangle, |y\rangle, |z\rangle$ via $|0\rangle=|z\rangle, \, |+1\rangle = -(|x\rangle +i |y\rangle)/\sqrt{2}, \ |-1\rangle\equiv (|x\rangle -i|y\rangle)/\sqrt{2}$.
From this, we see that all the Pauli gates $X$, $Y$, $Z$ and their linear combinations can be realized by suitable measurement bases.

Since the construction of the AKLT state is via singlets, the maximal spin magnitude of two sites cannot exceed unity and thus the AKLT state must be annihilated by the spin-2 projector on two neighboring sites. This leads to a parent Hamiltonian for the AKLT state
\begin{equation}
\label{eqn:HAKLT1D}
H=\frac{1}{2}\sum_{i}\left( \vec{S}_i\cdot \vec{S}_{i+1}+\frac{1}{3}  \big(\vec{S}_i\cdot \vec{S}_{i+1}
\big)^2+\frac{2}{3}\right).
\end{equation}
The Hamiltonian was shown to possess a unique ground, i.e., the AKLT state and to have a finite spectral gap above the ground state~\cite{AKLT,AKLT2,Knabe}. For two dimensions, establishing such a gap is highly nontrivial.

\begin{figure}[t]
   \includegraphics[width=0.5\textwidth]{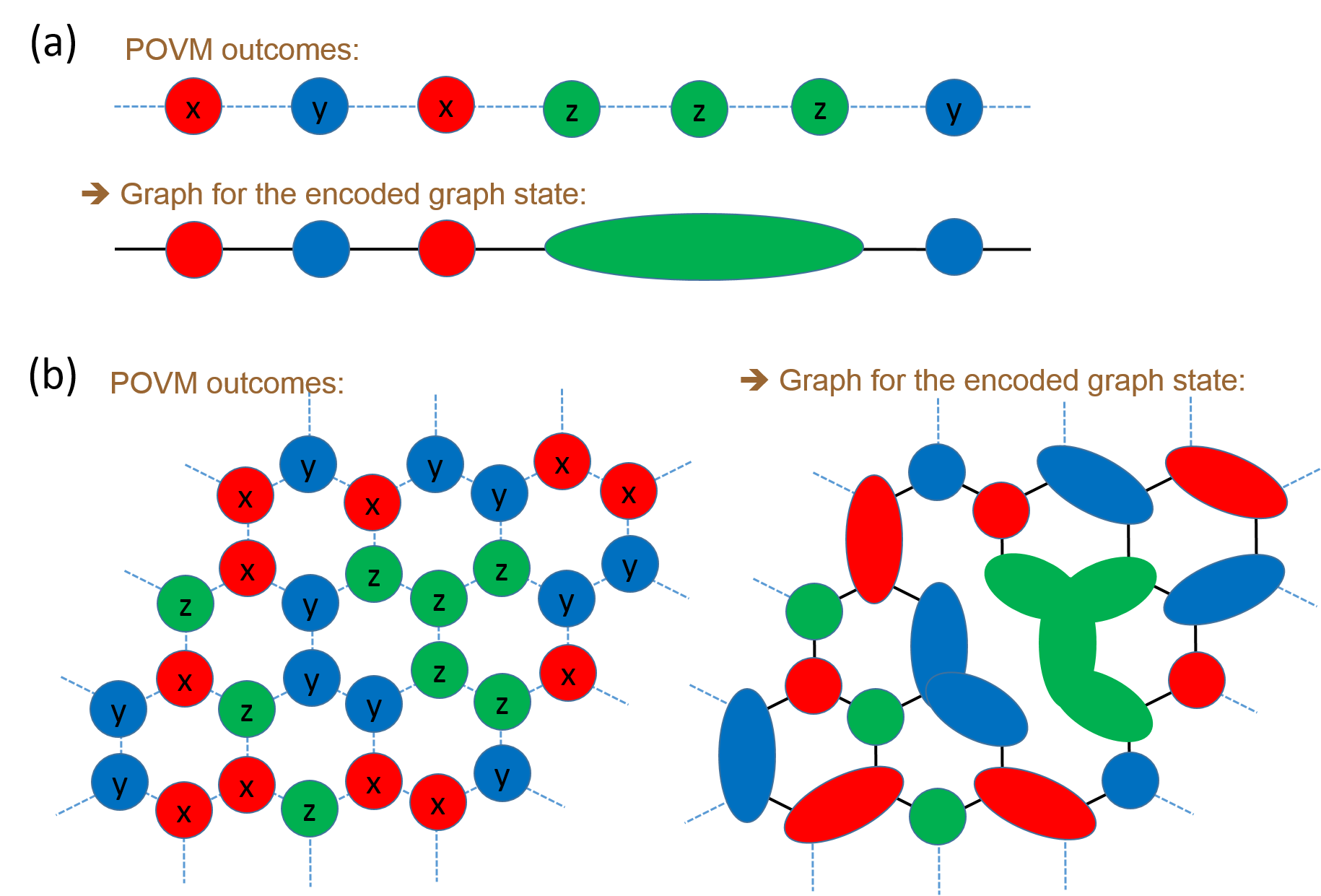}
  \caption{\label{fig:encoding}
  From AKLT to a graph state.  (a) A spin-1 chain; (b) Spin-3/2 case on the honeycomb lattice. Encoding is indicated by shapes with same color.  The new edges are derived from the original edges in a mod-2 fashion, as can be seen in (b). }
\end{figure}
\subsection{From AKLT to cluster state}

It turns out that by performing local measurements, the 1D AKLT state can be converted to the 1D cluster state, with reduced number of sites. The key point is that there is a
two-dimensional subspace spanned by $|S=1,S_z=1\rangle$ and
$|S=1,S_z=-1\rangle$ or equivalently by the two virtual qubits
$|00\rangle$ and $|11\rangle$, which also holds similarly for $S_x=\pm1$ and $S_y=\pm1$ subspaces.  One can therefore consider
\begin{eqnarray}
\!\!\!\!\!\!\!\!\!\!&&F_x = (\ketbra{S_x=1}+\ketbra{S_x=-1})/\sqrt{2}, \\
\!\!\!\!\!\!\!\!\!\!&&F_y=
(\ketbra{S_y=1}+\ketbra{S_y=-1})/\sqrt{2}, \\
\!\!\!\!\!\!\!\!\!\!&&F_z= (\ketbra{S_z=1}+\ketbra{S_z=-1})/\sqrt{2},
\end{eqnarray}
 as 
projections that preserve a two-dimensional subspace, where we
suppress the label $S=1$. Note that  the completeness relation in the spin-1 Hilbert space:
\begin{equation} 
\label{eqn:FF}
\sum_{\alpha=x,y,z}
F^\dagger_\alpha F_\alpha =\openone_{S=1},
\end{equation}
which means that a generalized measurement can be designed such that depending on the measurement outcome $\alpha=x,y, {\rm or}\, z$, a state $|\psi\rangle$ is taken to $F_\alpha|\psi\rangle$.

We note that the above $F$'s constitute the so-called generalized measurement or
POVM, characterized by $\{F_\alpha^\dagger F_\alpha\}$. Their
physical meaning is to define a two-dimensional subspace and to
specify a preferred quantization axis $x$, $y$ or $z$. In principle,
the POVM can be realized by a unitary transformation $U$ jointly on
a spin-1 state, denoted by $|\psi\rangle$, and a meter state
$|0\rangle_m$ such that
\begin{equation}
U |\psi\rangle|0\rangle_m = \sum_{\alpha}
F_\alpha|\psi\rangle|\alpha\rangle_m,
\end{equation}
where for the meter states $\langle
\alpha|\alpha'\rangle=\delta_{\alpha,\alpha'}$. A measurement on the
meter state will result in a random outcome $\alpha$, for which the
spin state is projected to
$F_\alpha|\psi\rangle$~\cite{NielsenChuang00}.

It was shown that  after performing the generalized
measurement on all sites, with $\{a_v\}$ denoting the measurement
outcomes ($x$, $y$ or $z$), the resulting state 
\begin{equation}
\label{eqn:1Dpost} |\psi(\{a_v\})\rangle\equiv \bigotimes_{v}
F_{v,a_v} |\Phi^{(1D)}_{\rm AKLT}\rangle
\end{equation}
is an ``encoded'' 1D cluster state~\cite{WeiAffleckRaussendorf11,WeiAffleckRaussendorf12}; see Fig.~\ref{fig:encoding}a. The encoding means that a logical cluster state qubit can effectively be represented by a few spin-1 sites, but, if desired, the encoding can be reduced to a single site by further local measurements. 

For example, for three consecutive sites with outcome $z$, the encoding is such that $|\uparrow\downarrow\uparrow\rangle$ and $|\downarrow\uparrow\downarrow\rangle$ represent the basis for a logical qubit, where $|\uparrow\rangle=|S=1,S_z=1\rangle$ and $|\downarrow\rangle=|S=1,S_z=-1\rangle$. Any state $a |\uparrow\downarrow\uparrow\rangle\otimes|\phi_0\rangle + b|\downarrow\uparrow\downarrow\rangle\otimes|\phi_1\rangle$ can be reduced to  $a |\uparrow\rangle\otimes|\phi_0\rangle + b|\downarrow\rangle\otimes|\phi_1\rangle$ (up to $\pm1$ relative phase dependent on measurement outcomes) by measuring two of the spins in the $(|\uparrow\rangle\pm\downarrow\rangle)/\sqrt{2}$ basis.

We wish to point that an alternative measurement scheme for such a conversion was first used by Chen et al.~\cite{ChenZeng}, but the POVM used here can be extended to two dimensions in the honeycomb and square lattices, allowing proof of universality of AKLT states on these lattices, as discussed below.
\subsection{Connection to SPT order}
Symmetry-protected topological (SPT) order~\cite{SPTRG,1DSPTcomplete,PollmannSPT,SPTtensor,ChenScience,cohomology} has been identified and characterized recently. But its origin dates back to Haldane's conjecture on the spectral gap of integer-spin antiferromagnetic Heisenberg chains, in particular, the spin-1 chain~\cite{Haldane,Haldane2}.
In the context of quantum computation, it was first revealed in the 1D SPT phase by Else et al.~\cite{ElseSchwarzBartlettDoherty}, where 
protection of certain quantum gates in MBQC arises due to the SPT order.  Examples include the 1D cluster state and the 1D spin-1 AKLT state that we discussed above~\cite{Gross,Gross2,Brennen}. Further exploration has been made and the connection is  further strengthened~\cite{MillerMiyake,Abhi,David1,David2}. 

Take the cluster state for example. For convenience, we consider group two neighboring spins into one physical site. Then we can consider four possible matrices $A[\pm\pm]$, where e.g. $A[++]=A[+]\cdot A[+]$ and $A[+]=(A[0]+A[1])/\sqrt{2}$. These matrices turn out to be $A[++]=\openone$, $A[+-]=Z$, $A[-+]=X$, $A[--]=XZ=-iY$. The cluster state and its Hamiltonian are invariant under global $Z_2\times Z_2$ symmetry generated by $X\otimes I$ and $I\otimes X$. Under the former, $A[\alpha,\beta]$ transforms as $Z\cdot A[\alpha,\beta]\cdot Z$, and under the latter $A[\alpha,\beta]\rightarrow X\cdot A[\alpha,\beta]\cdot X$. If one considers a long chain with two ends open, then the transformation on any one end become a projective representation of $Z_2\times Z_2$, given by the four Pauli matrices. The projectiveness of the representation is a signature of 1d SPT order. 

Similar conclusion is reached for the AKLT state if one considers the symmetry group generated by $\pi$ rotation around $x$ and $z$ spin axes.
Else et al.~\cite{ElseSchwarzBartlettDoherty} found that for the entire nontrivial SPT phase protected by $Z_2\times Z_2$ symmetry, the ground states are always represented by the MPS forms
$A[\alpha]=\sigma_\alpha \otimes B_\alpha$, where $\alpha$ denotes the physical indices, and $B_\alpha$'s are arbitrary matrices not constrained by the symmetry. If one can encode quantum information in the virtual subspace that the Pauli matrices $\sigma_\alpha$ act on, then perfect teleportation can be achieved across the entire spin chain. Such capability for teleportation has been generalized to other symmetry groups~\cite{Abhi,MillerMiyake}.

In particular, Miller and Miyake used a different symmetry group, i.e., $S_4$, and showed that beyond just the teleportation, arbitrary one-qubit gates can be implemented in the entire phase~\cite{MillerMiyake}. Such universality for qubit or qudit gates has been greatly generalized and the connection of SPT order and quantum computation is firmly strengthened in 1d~\cite{David1,David2}.
\section{Two dimensions}
\subsection{Cluster state on square and other lattices}
When a graph state is defined on a regular lattice, such as the 1d array, the square, honeycomb, or triangular lattice, we will refer to this state as a cluster state. Briegel and Raussendorf first invented the cluster state on the square lattice and characterized its entanglement properties~\cite{BriegelRaussendorf}. Soon after Raussendorf and Briegel showed that this cluster state can be used to simulate universal gates and thus implement universal quantum computation~\cite{Oneway}. First we will illustrate that cluster states on 2d regular lattices are interconvertible by using only local Pauli measurements, first shown by Van den Nest et al.  that in addition to the square lattice, cluster states on other regular lattices can also be used for universal quantum computation~\cite{VandenNest1}.  They explicitly demonstrated that by using only local Pauli Z and Y measurements, the underlying graphs characterizing the cluster states can be converted from the hexagonal lattice, to the triangular lattice, to the kagome lattice, then to the square lattice, demonstrating these other cluster states are also universal.

Instead of using the square-lattice cluster state, we will illustrate that universal gates can be realized on a specific brickwork lattice. The cluster state on the brickwork also offers an interesting type of computation: the blind quantum computation, where computation can be done without revealing what was computed~\cite{BlindQC,BlindQCScience}. 

\subsection{The brickwork lattice and universality}
\begin{figure}
   \includegraphics[width=0.48\textwidth]{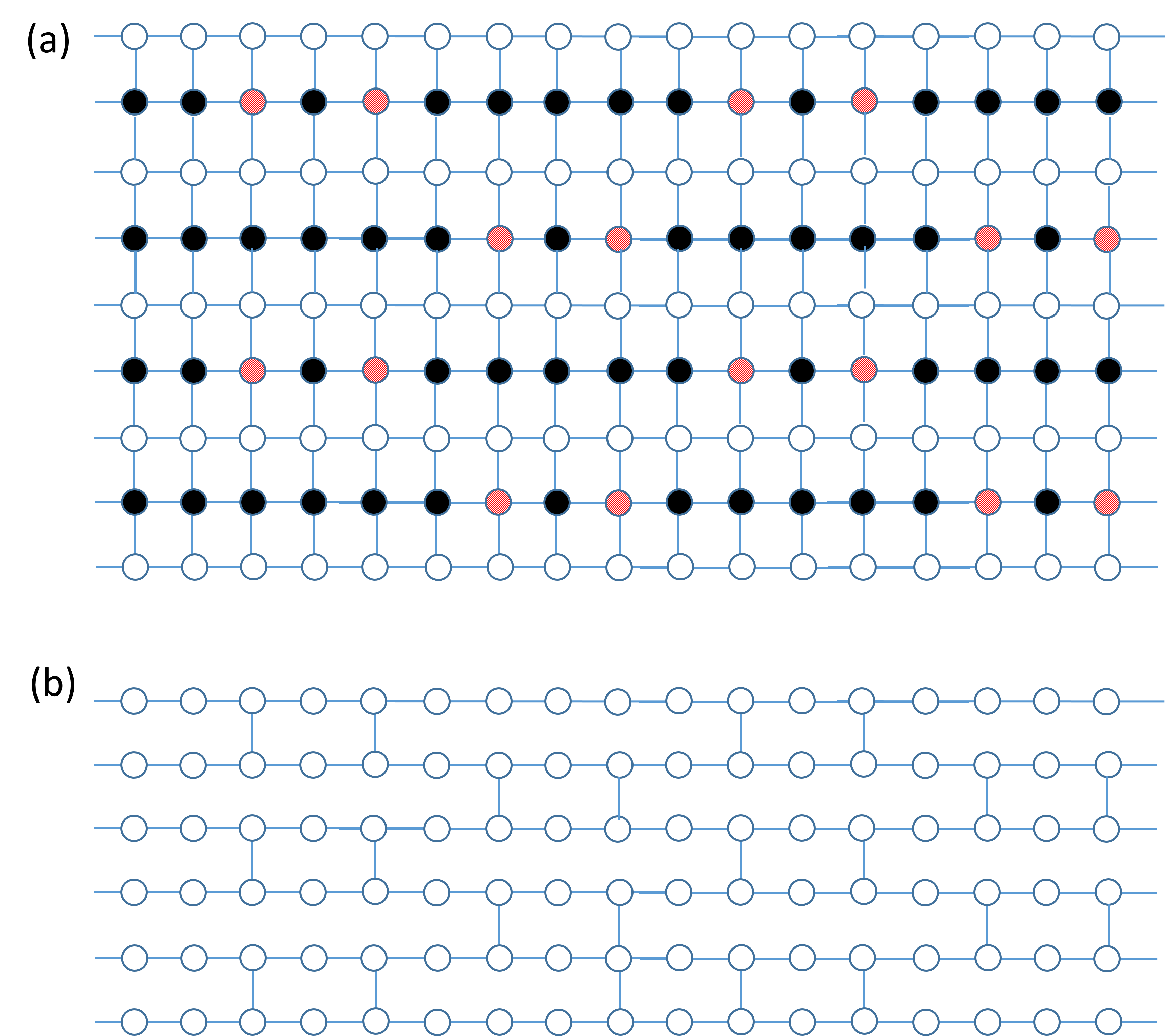}
  \caption{\label{fig:brickwork}
  Conversion of graph states: (a) Square lattice; (b) Brickwork lattice. Solid black circles indicate $Z$ measurements and meshed red circles indicate $Y$ measurement. By performing these measurements, the graph state on the square lattice will be converted to that on the brickwork lattice.  }
\end{figure}
We show the brickwork lattice in Fig.~\ref{fig:brickwork} and demonstrate how to convert from the square lattice cluster state to the brickwork state. Then we will use the brickwork state for explaining universal measurement-based quantum computation.
The brickwork state was initially used in demonstrating blind quantum computation by Broadbent, Fitzsimons and Kashefi~\cite{BlindQC}.
\begin{figure}
   \includegraphics[width=0.4\textwidth]{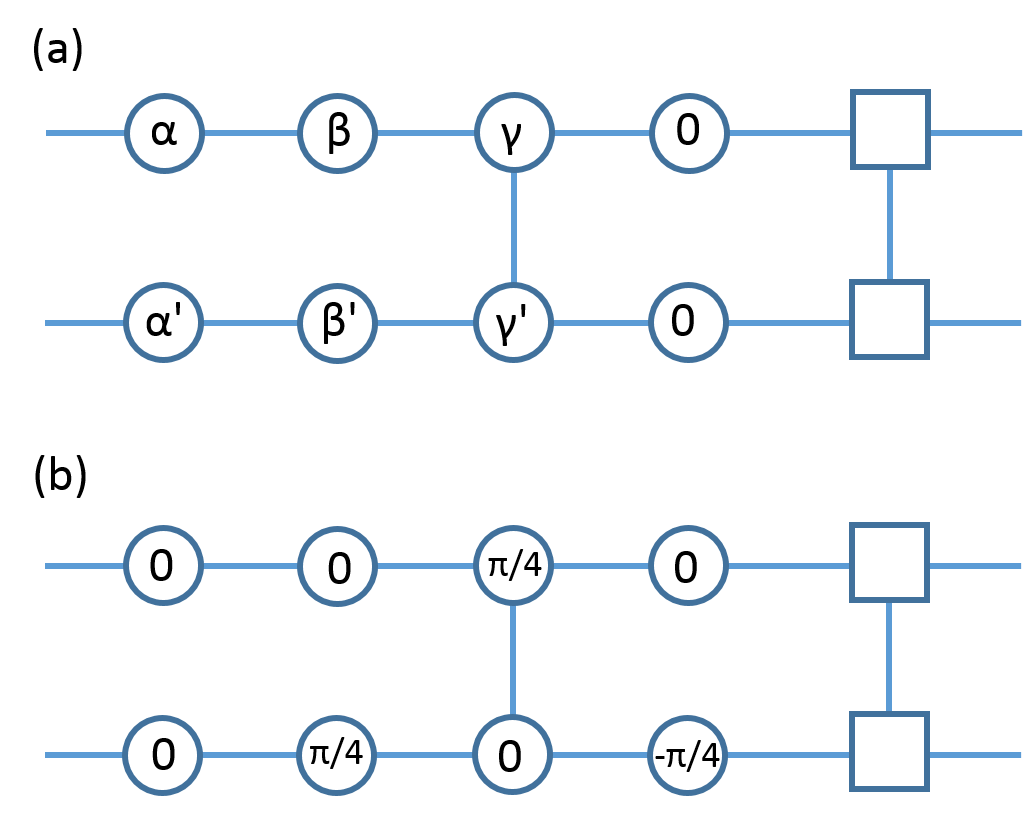}
  \caption{\label{fig:gates}
Gates implementation: (a) two indepedent one-qubit gates; (b) CNOT gate. The symbols inside the circles indicate the measurement angles.}
\end{figure}
The construction of universal gates, including arbitrary single-qubit gates and a CNOT gate, is illustrated in Fig.~\ref{fig:gates}.  The proof proceeds in the way similar to the arbitrary qubit gates implemented on a segment of the 1d cluster state in Sec.~\ref{sec:1dcluster}, except the two CZ gates present between two wire segments need to be inserted and taken into account.  For example,  the single-qubit gates can be proved by translating the circuit in Fig.~\ref{fig:gates}a to the following expression
\begin{eqnarray}
U&=&{\rm CZ} ( HZ^{s})\otimes 
   (  HZ^{s'}) (H e^{i\gamma Z/2} Z^{s_\gamma})\otimes 
   ( e^{i\gamma' Z/2} Z^{s_{\gamma'}}) \nonumber \\
   &&{\rm CZ} (H e^{i\beta Z/2} X^{s_\beta})\otimes 
   (H e^{i\beta' Z/2} Z^{s_{\beta'}})\nonumber \\
  && (H e^{i\alpha Z/2} Z^{s_\alpha})\otimes 
   ( H e^{i\alpha' Z/2} Z^{s_{\alpha'}}), \label{eqn:UU}
\end{eqnarray}
which is a direct translation from the circuit in Fig.~\ref{fig:gates}a. The CNOT gate can be proved by translating the circuit in Fig.~\ref{fig:gates}b to
\begin{eqnarray}
U&=& {\rm CZ} (H Z^{s_4})\otimes 
   (H e^{-i\pi Z/4} Z^{s_{4'}}) 
   (H e^{i\pi Z/4} Z^{s_3})\otimes 
   (H Z^{s_{3'}}) \nonumber \\
   && {\rm CZ} (H  Z^{s_2})\otimes 
   (H e^{i\pi Z/4} Z^{s_{2'}})
   (H  Z^{s_1})\otimes 
   (H  Z^{s_{1'}}). \label{eqn:CNOT}
\end{eqnarray}
By placing all these gates appropriately on the brickwork lattice, universal quantum computation can be achieved.  Since all 2d graph states on regular lattices can be converted by local measurements to the square lattice, then they are all universal resources.  

The byproduct operators are single-qubit operators and can be propagated forward or incorporated by modifying later measurement basis. 
\subsection{Random planar graph states}
To go beyond the regular lattices, Browne et al. considered faulty square lattices, namely, each lattice site is occupied with a probability $p$ (in the sense of site percolation)~\cite{Browne}. They showed that as long as $p$ is above the site percolation threshold $p_c$ of the square lattice, then the graph state on the faulty square lattice is still universal. The key idea is that there is still sufficient connectivity in the graph such that a honeycomb lattice (as a subgraph) can be distilled out by using local measurements. This subsequently led to the general connection of percolation to universality of random planar graph states~\cite{Browne,WeiAffleckRaussendorf12}, whose graphs reside in the supercritical phase of percolation. 
\subsection{Nielsen's no-go theorem}
Nielsen proved that the cluster state cannot appear as the exact ground state of any ``naturally occurring physical system'', by which he means Hamiltonians with two-body interactions~\cite{Nielsen}. If it could, then cooling the system could lead to creation of the cluster state and subsequent local measurements would enable universal quantum computation.  Nielsen's proof uses the so-called stabilizer formalism and the idea from quantum error correction. Consider the Hamiltonian $H=\sum_{\sigma,\tau} h_{\sigma,\tau}\sigma\otimes \tau$, where $\sigma$ and $\tau$ are taken from either the identity or the three Pauli operators.  Since for any vertex $u$, the cluster state has the stabilizer operator ${\cal K}_u$ as in Eq.~(\ref{eqn:Ku}), one considers
\begin{equation}
(\sigma\otimes\tau) {\cal K}_u (\sigma\otimes\tau) = n_u(\sigma,\tau) {\cal K}_u, 
\end{equation}
where $n_u(\sigma,\tau)=\pm 1$ represents the syndrome caused by applying the error operator $\sigma\otimes\tau$ to the cluster state. We denote $\vec{n}(\sigma,\tau)$ with $[\vec{n}]_u=n_u(\sigma,\tau)$  as the syndrome vector corresponding to $\sigma\otimes\tau$. If for different combination of $(\sigma,\tau)$, their syndrome vectors are all distinct, then the states $\sigma\otimes\tau |{\cal C}\rangle$ are orthonormal and the cluster state $|{\cal C}\rangle$ cannot be an eigenstate of the Hamiltonian $H$~\cite{Nielsen,Haselgrove}.

Chen et al. used a different approach and gave a general proof for such a no-go theorem for any qubit two-body frustration-free Hamiltonian~\cite{ChenChenDuanJiZeng}. The proof uses the equivalence of quantum states under stochastic
local operation and classical communication and the result by Bravyi on the homogeneous Hamiltonians~\cite{Bravyi}. This was discussed in a recent review~\cite{Kwek}, so we will not repeat the details here.  Ji, Wei and Zeng further characterized the ground space of two-body frustration-free Hamiltonians~\cite{JiWeiZeng}. 

We remark that the above no-go theorem applies only to the exact ground state. If we allow approximate ground state, there can be such a two-body Hamiltonian, according to a perturbative construction by Rudolph and Bartlett~\cite{RudolphBartlett}. Movitated by these no-go results, Chen et al. constructed a spin-5/2 state, the so-called Tri-Cluster state~\cite{Tricluster}, such that (i) it is the unique ground state of a two-body Hamiltonian and (ii) it is a universal resource for MBQC. The quest for other two-body interacting ground states followed suit. So far, the universal resource states that are the unqiue ground state of two-body interacting Hamiltonians have the smallest local Hilbert-space dimension being 4 (i.e. spin-3/2), such as the spin-3/2 AKLT state. Whether a universal resource state is of spin-1 entity and the unique ground state of a two-body interacting Hamiltonian remains an open question.
 
\subsection{2d AKLT states}
\begin{figure}
   \includegraphics[width=0.48\textwidth]{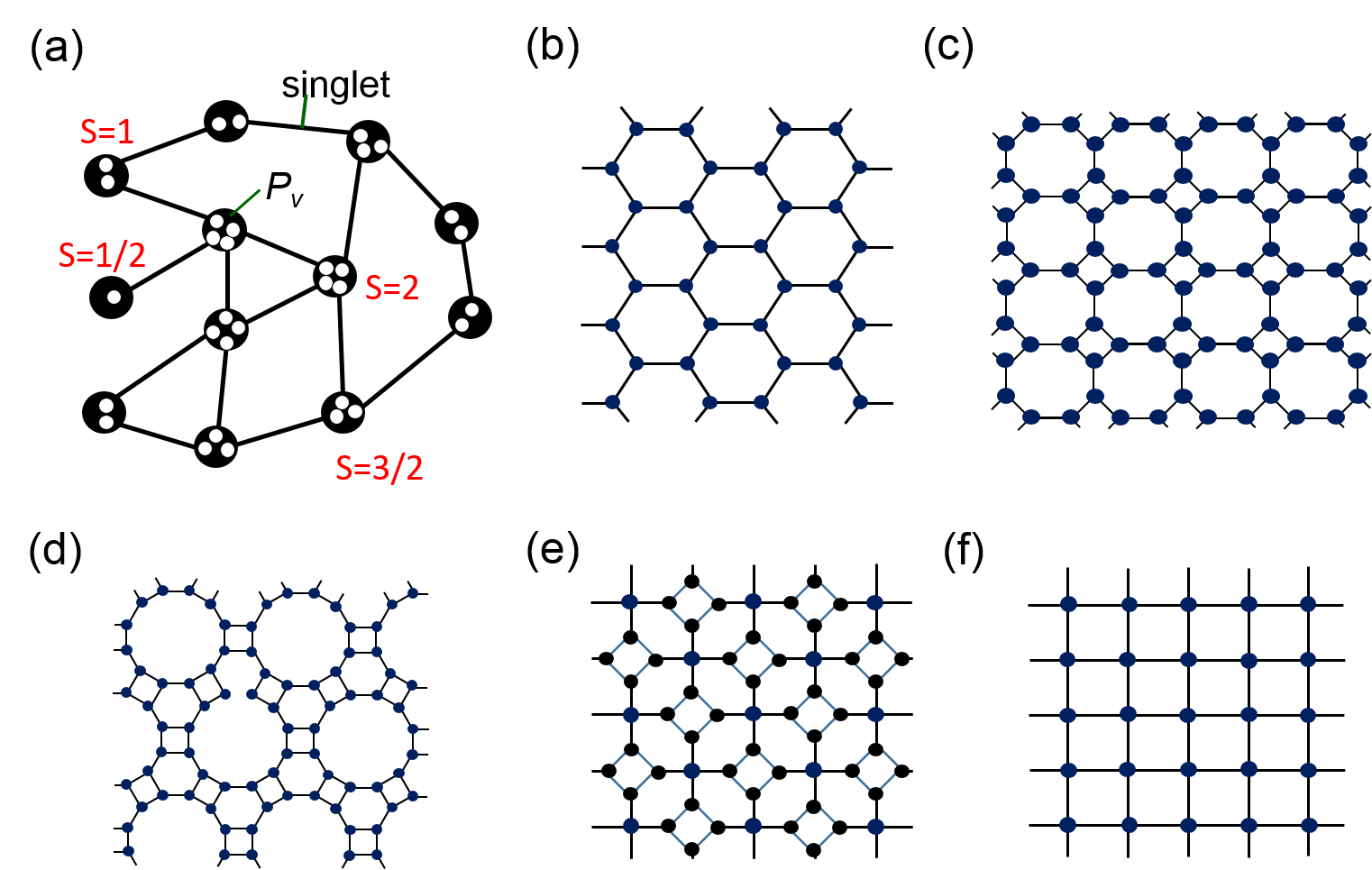}
  \caption{\label{fig:akltGraphs}
  AKLT states on graphs or 2d lattices. (a) Arbitrary graph: showing virtual qubits on each site. The physical spin is a projection $P_v$ from the symmetric subspace of the virtual qubits to a spin $S$ Hilbert space, where $S$ is half the number of virtual qubits or equivalently of neighbors. (b) Honeycomb lattice. (c) Square-octagon lattice. (d) Cross lattice. (e) Hexagon-square lattice. (f) Square lattice.  }
\end{figure}
Since the 1d AKLT state and later its generalizations have been successfully shown to enable arbitrary qubit or qudit gates, it is natural to ask whether it also works for 2D AKLT states, especially due to their uniqueness as ground states of two-body AKLT Hamiltonians. One and two-dimensional AKLT states are characterized as some kind of disordered states, having no local magnetizations and no breaking of lattice translation, and their construction is via the valence-bond approach, described above. Although they were also referred to as valence-bond-solid states, to avoid confusion from the most common known valence-bond solids~\cite{VBS}, which spontaneously break lattice translation and rotation symmetry, we will just refer to these AKLT states as valence-bond states. However, AKLT states on three dimensions need not be disordered, they can possess N\'eel order, depending on the lattice~\cite{ParameswaranSondhiArovas09}.  Even spin glass can appear as ground states of AKLT models on random graphs~\cite{spinglass}. Ordered states, such as ferromagnetic or N\'eel states, are conjectured not to possess sufficient entanglement structure for universal MBQC.

The first member of 2d AKLT family shown to be a universal resource is the spin-3/2 state on the honecomb lattice~\cite{WeiAffleckRaussendorf11,Miyake11}. Subsequent works have led to a better understanding of the universality~\cite{Wei13,WeiPoyaRaussendorf,WeiRaussendorf15}, albeit incomplete. What is known is as follows.  AKLT states involving spin-2 and other lower spin entities are universal if they reside on a 2D frustration-free regular lattice with any combination of spin-2, spin-3/2, spin-1 and spin-1/2 (consistent with the lattice structure). Here ``lattice'' is used loosely to denote a 2d periodic titling which can have boundary.  However, how do we go beyond spin-2 and determine whether other higher-spin AKLT states are also universal  remains an open question. One obstacle is the lack of a suitable POVM or any other local measurement to show that universal gates can be implemented. Could there be some kind of ``order parameter'' for quantum computational universality, at least in certain family of states?

The spin-3/2 AKLT state on any trivalent lattice is a ground state of the following Hamiltonian
\begin{equation}
\label{eqn:HAKLT32}
 H_{AKLT}^{S=3/2}=\!\!\sum_{{\rm edge}\,\langle i,j\rangle}\Big[ \vec{S}_i\cdot \vec{S}_{j}+\frac{116}{243}(\vec{S}_i\cdot \vec{S}_{j})^2+\frac{16}{243}(\vec{S}_i\cdot
 \vec{S}_{j})^3\Big],
\end{equation}
where an irrelevant constant term has been dropped. To show that it is a universal resource, a suitable POVM (for $S=3/2$) is as follows:
\begin{eqnarray}
\!\!\!\!\!\!\!\!\!\!&&F_x = \frac{\sqrt{2}}{\sqrt{3}}(\ketbra{S_x=\frac{3}{2}}+\ketbra{S_x=-\frac{3}{2}}), \\
\!\!\!\!\!\!\!\!\!\!&&F_y=
\frac{\sqrt{2}}{\sqrt{3}}(\ketbra{S_y=\frac{3}{2}}+\ketbra{S_y=-\frac{3}{2}}), \\
\!\!\!\!\!\!\!\!\!\!&&F_z=\frac{\sqrt{2}}{\sqrt{3}} (\ketbra{S_z=\frac{3}{2}}+\ketbra{S_z=-\frac{3}{2}}).
\end{eqnarray}
  Note that the prefactor $\sqrt{2}/\sqrt{3}$ differs from that of $1/\sqrt{2}$ in the $S=1$ case and one can verify the completeness relation in the spin-3/2 Hilbert space:
\begin{equation} 
\label{eqn:FF32}
\sum_{\alpha=x,y,z}
F^\dagger_\alpha F_\alpha =\openone_{S=3/2},
\end{equation}
which means that a generalized measurement can be designed such that depending on the measurement outcome $\alpha=x,y, {\rm or}\, z$, a state $|\psi\rangle$ is taken to $F_\alpha|\psi\rangle$.

Similarly to the 1d case, one can show that regardless of the outcomes  $\{a_v\}$, the post-POVM state 
\begin{equation}
\label{eqn:2Dpost} |\psi(\{a_v\})\rangle\equiv \bigotimes_{v}
F_{v,a_v} |\Phi^{(S=3/2)}_{\rm AKLT}\rangle
\end{equation}
is an ``encoded'' graph state, whose graph is modified from the original trivalent lattice, e.g., the honeycomb, in a way determined by $\{a_v\}$.  The encoding means that a logical cluster state qubit can effectively be represented by a few spin-3/2 sites, but, if desired, the encoding can be reduced to a single site by further local measurements; see also Fig.~\ref{fig:encoding}b. A sufficient condition for quantum computational universality is thus to check to see if the resulting random graphs are, with nonzero probability, in the supercritical phase of percolation.

The POVM cannot be trivially extended to $S=2$ case, but it happens that in this case one only needs a second round of measurements to obtain 2d planar graph states. The extension to higher spins for the proof of universality is current unknown.

Nevertheless we note that Eq.~(\ref{eqn:2Dpost}) will give a graph state even for any spin-$S$ AKLT state,  where
\begin{equation} 
\label{eqn:generalF}
F_\alpha\sim |S_\alpha=S\rangle\langle S_\alpha=S|+|S_\alpha=-S\rangle\langle S_\alpha=-S|.
\end{equation} 
However,
for $S=2$ and higher $S$,  $\sum_{\alpha=x,y,z}
F^\dagger_\alpha F_\alpha$ is not proportional to the identity operator. But  for $S=2$ one only needs a second round of measurements to obtain 2D planar graph states. It requires a few more technicalities, which we do not have space to discuss  here, to demonstrate that the spin-2 AKLT state on the square lattice (as well as on the diamond lattice) is a universal resource~\cite{WeiRaussendorf15}.

Most of the discussions on AKLT states naturally use the singlet state $|01\rangle-|10\rangle$ as the valence bond. However, one may as well use other maximally entangled states, such as $|00\rangle\pm|11\rangle$  or $|01\rangle+|10\rangle$. For bipartice lattices, these AKLT states can be inter-converted locally, so the quantum computational capability is indentical~\cite{HuangWagnerWei}. But on non-bipartite lattices they cannot be inter-converted. AKLT states using different bonds on these lattices will may different capability for quantum computations. This has not been much explored.  

It was known that the AKLT states on both the honeycomb and square lattices possess exponentially decaying correlation functions~\cite{AKLT2}. Even though these are strong evidence for the finite spectral gap, so far no rigorous proof has been given. Recent developments in tensor network methods have allowed estimation of these gaps in the thermodynamic limits, and strong numerical evidence supports the finiteness of the spectral gaps~\cite{AKLTgap,VerstraeteGroup}.

\subsection{Away from the AKLT point}

Niggemann, Kl\"umper and Zittarz considered deformation on AKLT states and found that there is additional a N\'eel phase under certain deformation, and the transition is of Ising like~\cite{NiggemannKlumperZittarz}. The scenario also occurs at other lattices, such as the square-octagon and the square lattices~\cite{NiggemannZittartz,NKZspin2}.  
Darmawan, Brennen and Barlett considered the one-parameter $a$ family of the deformed spin-3/2 AKLT state on the honeycomb lattice. The Hamiltonian can be written as
\begin{equation}
H_{\rm NKZ}={\sum}_{\langle i,j\rangle} [D(a)_i\otimes
D(a)_j]h_{i,j}^{\rm AKLT} [D(a)_i\otimes D(a)_j]^\dag\,,
\end{equation}
where
$h_{i,j}^{\rm AKLT}$ is the two-body term in the AKLT Hamiltonian~(\ref{eqn:HAKLT32}) and the deformation
operator is defined as $D(a)=\mbox{diag}(\sqrt{3}/a,1,1,\sqrt{3}/a)$
in the spin $S_z$ basis, where $a>0$. The ground state is, by construction of the Hamiltonian,
\begin{equation}
\ket{\psi(a)_{\rm
NKZ}}\propto(D(a)^{-1})^{\otimes N}\ket{\psi_{\rm AKLT}}.
\end{equation}

There were able to construct a modified POVM (parameter-dependent) such that the action is effectively to (i) first locally undo the local deformation $D(a)^{-1}$, and (ii) second, perform the original POVM for the AKLT state. They showed that for a certain range of the parameter $a$, the universality still persists. Interestingly the termination of the quantum computational power coincides with the transition to the N\'eel ordered phase~\cite{DarmawanBrennenBartlett}. This was shown to be the case for other trivalent lattices~\cite{HuangWagnerWei}. However, for the parameter $a$ being small ($a<1$), due to some technicality the quantum compuational power has not been explored.


Using the result in Ref.~\cite{WeiAffleckRaussendorf11} that the POVM on all sites converts the AKLT state to a graph state, whose graph depends on the patterns,
\begin{equation}
|{\cal C}_{\rm pattern}\rangle=\mathop{\otimes}_{\alpha_v={\rm pattern}} F_{\alpha_v}|\psi_{\rm AKLT}\rangle,
\end{equation}
Darmawan and Bartlett directly considered the pattern defined by $\{\alpha_v\}$ and the projectors $P_{\alpha_v}$ associated with $F_{\alpha_v}$ (normalizing $F$ such that $P^2=P$) and constructed a one parameter $\delta$ family of deformed wave function~\cite{Darmawan2}
\begin{equation}
|\psi_{\{\alpha_v\};\delta}\rangle\equiv  \mathop{\otimes}_v D_v (\delta) |\psi_{\rm AKLT}\rangle,
\end{equation}
where $D_v (\delta)\equiv (1-\delta)  P_{\alpha_v} + \delta\openone_v$.
They showed that the previously known non-universal AKLT state on the star lattice can be deformed to one that becomes universal, with an appropriate pattern $\{\alpha_v\}$. Moreover, the resultant parent Hamiltonian is shown to be gapped if $\delta_c>\delta>0$ for some nonzero $\delta_c$~\cite{DarmawanBartlett16}. This work makes some progress towards proving the spectral of the Hamiltonian.
\subsection{Fixed-point SPT states}
\begin{figure}
   \includegraphics[width=0.48\textwidth]{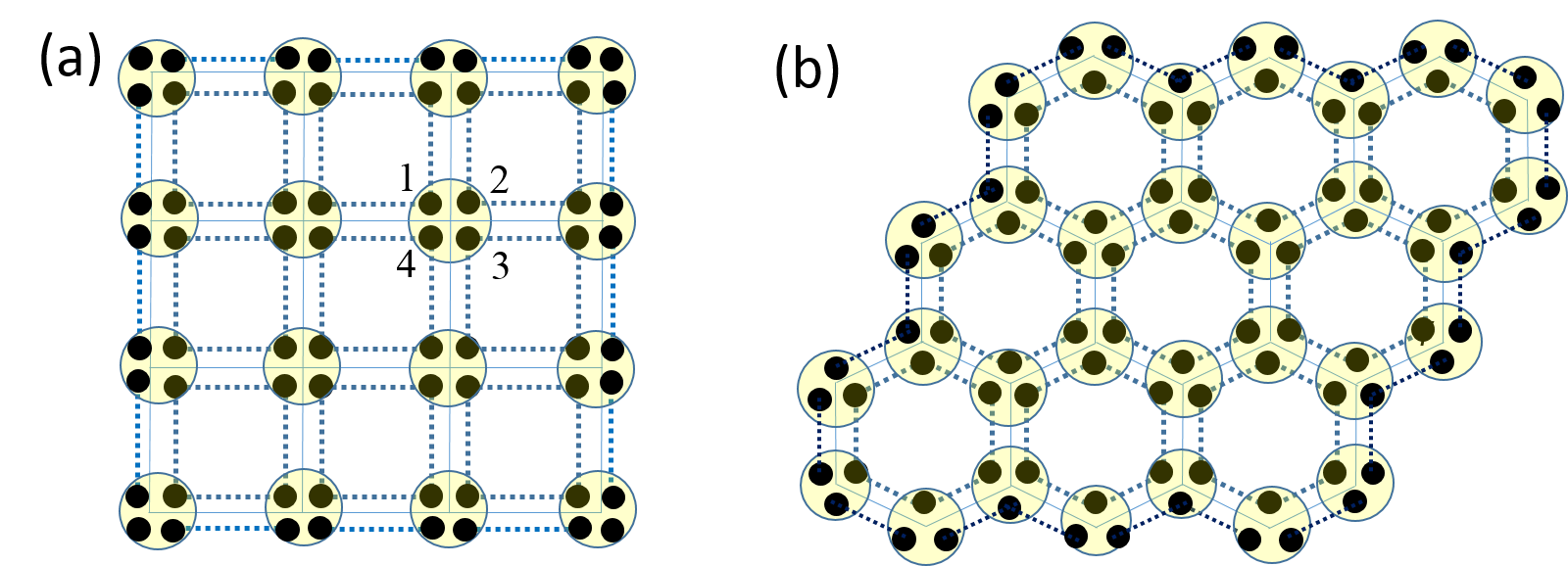}
  \caption{\label{fig:ghz}
  Fixed-point SPT states: (a) on the square lattice; (b) on the honeycomb lattice. }
\end{figure}
\begin{figure}
   \includegraphics[width=0.48\textwidth]{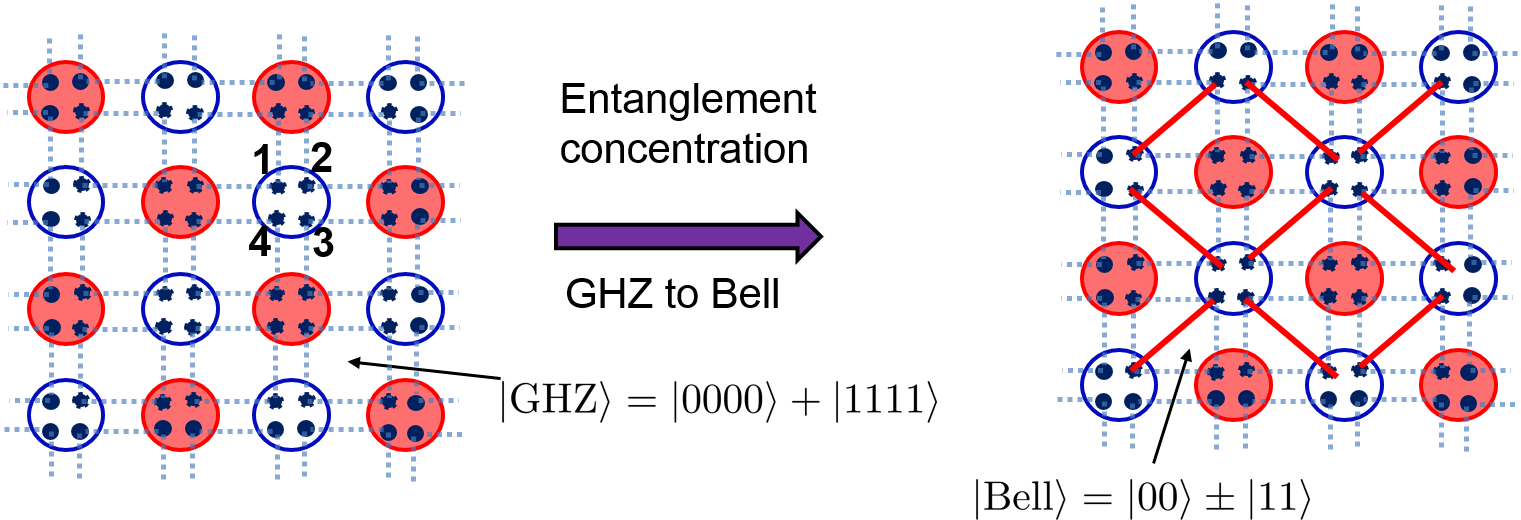}
  \caption{\label{fig:entanglementConcentration}
  Entanglement concentration: converting the SPT state to a known universal resource state.}
\end{figure}

The first discovered 2d bosonic SPT order is the CZX model~\cite{CZX}. In this model, each site contains four qubits or ``partons'' and each qubit forms with corresponding ones on neighboring sites in the  GHZ states $|0000\rangle+|1111\rangle$. The ground state is essentially composed of product of such GHZ states; see illustrations in Fig.~\ref{fig:ghz}. The on-site nontrivial $Z_2$ action is the so-called $U_{\rm CZX}=U_{\rm CZ}U_X$, where $U_X=X_1\otimes X_2\otimes X_3\otimes X_4$ is a product of Pauli X operators on all qubits in a site (with the numbers labeled in a clockwise way) and the $U_{\rm CZ}={\rm CZ}_{12} {\rm CZ}_{23} {\rm CZ}_{34} {\rm CZ}_{41}$; see also Fig.~\ref{fig:entanglementConcentration}. The 2d SPT order can be seen from the symmetry action at a boundary, whose degrees of freedom reside on the half GHZ plaquettes, cut by the boundary line. It is non-onsite and there are nontrivial 3-cocycles emerging from composing a sequence of three symmetry actions in different ways.
 
We generalize the constructions by Chen et al.~\cite{cohomology} to arbitrary 2d lattices and 2d planar graphs and showed that these GHZ-like plaquettes still exhibit nontrivial SPT order with appropriate symmetry actions defined via 3-cocycles~\cite{Hendrik}. Interestingly these states can serve as universal resource for MBQC, provided the graphs are in the supercritical phase of percolation (and regular 2d lattices certainly are).

\smallskip\noindent {\bf Nontrivial SPT order.}
To understand the nontrivial SPT order, we need to define an appropriate symmetry. To allow generality, the GHZ-like  maximally entangled state on a plaquette $p$ is 
\begin{equation}
|\psi\rangle_{p}=\frac{1}{\sqrt{|G|}}\sum_{g\in G} |\alpha_1=g,\alpha_2=g,\dots,\alpha_{k^*}=g\rangle,
\end{equation}
where the $|G|$ is the order of the group $G$ and $k^*$ is the number of ``partons'' within a plaquette. 
 We do not require $k^*$ to be constant across arbitrary lattices. We show that the product these plaquette states $|\psi_{\rm SPT}\rangle=\otimes_p |\psi\rangle_p$ is a nontrivial SPT state, with respect to onsite symmetry action defined below. Such states are illustrated in Fig.~\ref{fig:ghz} and the one on the square lattice was first investigated in the so-called CZX model by Chen et al.~\cite{CZX}.

The on-site representation of the symmetry action $g\in G$, $U^I_{i}(g)$, on site $i$ within the `sublattice' $I$ is then given by
\begin{align}
& U_i^I(g)\ket{\alpha_1,\alpha_2,...,\alpha_{k^*}}\nonumber\\
&=f^I_3(\alpha_1,\alpha_2,...,\alpha_{k^*}, g, \bar{g})\ket{g\alpha_1,g\alpha_2,...,g\alpha_{k^*}}\label{U}
\end{align}
with $k^*$ being the number of partons within one physical site, $\bar{g}$ being a fixed element in $G$ and 
\begin{align}
& f^I_3(\alpha_1,\alpha_2,...,\alpha_{k^*}, g, \bar{g})\nonumber\\
&\equiv\prod_{\{i_a\},\{i_b\}}\frac{\nu_3(\alpha_{i_a},\alpha_{i_a+1},g^{-1}\bar{g},\bar{g})}{\nu_3(\alpha_{i_b+1},\alpha_{i_b},g^{-1}\bar{g},\bar{g})}\label{f}.
\end{align}
The sublattices $I$  can be chosen arbitrarily, but for regular lattices, the colors in the colorability of graphs are such a choice. For simplicity we label partons $i$ so that $i\rightarrow i+1$ is counting counterclockwise, and $\alpha_{k^*+1}\equiv \alpha_1$ and $\alpha_{0}\equiv \alpha_{k^*}$. The two non-overlapping sets ($\{i_a\}\cap\{i_b\}=\emptyset$ but  $\{i_a\}\cup\{i_b\}=\{1,...,k^*\}$) of indices $\{i_a\},\{i_b\}\subset\{1,...,k^*\}$ define the branching structure of $f^I_3$; $\{i_a\}$ gives counterclockwise branched tetrahedrons and  $\{i_b\}$ defines clockwise orientated tetrahedrons. $\nu_3(g_0,g_1,g_2,g_3)$ is a 3-cocycle.

To verify that
\begin{align}
U_i^I(g'g^{-1})U_i^I(g)\ket{\alpha_1,\alpha_2,...,\alpha_{k*}}=U_i^I(g')\ket{\alpha_1,\alpha_2,...,\alpha_{k*}},\label{linear}
\end{align}
we need compare the phase factors on both sides. The left side of Eq.~(\ref{linear}) gives
\begin{align}
&\prod_{\{i_a\},\{i_b\}}\frac{\nu_3(\alpha_{i_a},\alpha_{i_a+1},g^{-1}\bar{g},\bar{g})}{\nu_3(\alpha_{i_b+1},\alpha_{i_b},g^{-1}\bar{g},\bar{g})}\frac{\nu_3(g\alpha_{i_a},g\alpha_{i_a+1},gg'^{-1}\bar{g},\bar{g})}{\nu_3(g\alpha_{i_b+1},g\alpha_{i_b},gg'^{-1}\bar{g},\bar{g})},
\end{align}
which, by using  $g\cdot \nu_3(g_0,g_1,g_3)=\nu_3(gg_0,gg_1,gg_3)$, can be rewritten as
\begin{align}
&\prod_{\{i_a\},\{i_b\}}\frac{\nu_3(\alpha_{i_a},\alpha_{i_a+1},g^{-1}\bar{g},\bar{g})\nu_3(\alpha_{i_a},\alpha_{i_a+1},g'^{-1}\bar{g},g^{-1}\bar{g})}{\nu_3(\alpha_{i_b+1},\alpha_{i_b},g^{-1}\bar{g},\bar{g})\nu_3(\alpha_{i_b+1},\alpha_{i_b},g'^{-1}\bar{g},g^{-1}\bar{g})}.\label{phasecompare1}
\end{align}
The phase factor generated on the right side of Eq.~(\ref{linear}) gives 
\begin{align}
\prod_{\{i_a\},\{i_b\}}\frac{\nu_3(\alpha_{i_a},\alpha_{i_a+1},g'^{-1}\bar{g},\bar{g})}{\nu_3(\alpha_{i_b+1},\alpha_{i_b},g'^{-1}\bar{g},\bar{g})},\label{phasecompare2}
\end{align}
which equals to that on the left side. One can go on to show that $U(g)\equiv\mathop{\otimes}_{\{i,I\}} U_i^I(g)$ is indeed a global on-site symmetry.
This shows that the  state $|\psi_{\rm SPT}\rangle$ is a SPT state w.r.t. to the on-site symmetry $U(g)$, and if the 3-cocycle is nontrivial, then the SPT order is nontrivial.

\smallskip\noindent {\bf Quantum computational universality}.
One way to see how the universality can arise is to use the notion of entanglement concentration from GHZ states to Bell states. (i) For instance, an $n$-qubit GHZ state: $|{\rm GHZ}_n\rangle=|00\dots0\rangle + |11\dots 1\rangle$ after being measured on qubit 1 in the X basis, the remaining state will become a $(n-1)$-qubit GHZ state $|{\rm GHZ}_{n-1}^\pm\rangle=|00\dots0\rangle \pm |11\dots 1\rangle$, where $\pm$ sign depends on the outcome of X measurement. One can continue and concentrate the entanglement to a Bell state $|00\rangle+|11\rangle$ between any two qubits. The measurements are local. (ii) Now if a site contains two qubits that are respectively part of a GHZ state, then by local measurement on this site we can merge the two GHZ states into one GHZ state. All of these can be extended to the general qudit case. Graphically, we can draw one and only one line between any two qubits in a GHZ plaquette to represent the case (i) and can also draw one line cross neighboring plaquettes to link any of the two qubits on these plaquettes, using (ii).  One can use these to show that as long as the underlying graph for the  state  $|\psi_{\rm SPT}\rangle$ is in the supercritical phase of percolation, then it is a universal resource for MBQC~\cite{Hendrik}. As illustrated in Fig.~\ref{fig:entanglementConcentration}, the CZX SPT state is converted to a valence-bond state on a square lattice by local measurements.

Percolation arises due to the requirement to have sufficient or macroscopic number of universal gates for universal quantum computation. One drawback of using these fixed-point wave functions for MBQC is that the dimension of local Hilbert space is not small, e.g. 16 for the CZX state or 8 for the extension to the honeycomb lattice. A different construction by Miller and Miyake achieves qubit SPT states for universal MBQC~\cite{MillerMiyake16}, which will be discussed below. 

 It is also interesting to know whether the quantum computational universality is only a property of SPT fixed points or is a more general property. We have shown that by deforming these SPT wavefunctions away from the fixed points  there is a region in the single-parameter family that quantum computational universality persists~\cite{WeiHuang}. Moreover, the disapperance of this universality in one direction seems to coincidence with the phase transition from the SPT phase to a $Z_2$ symmetry-breaking one. What would be a potential breakthrough is to establish an entire SPT phase supporting universal MBQC, but this is still an open question. 
\subsection{Miller-Miyake states}
\begin{figure}
   \includegraphics[width=0.4\textwidth]{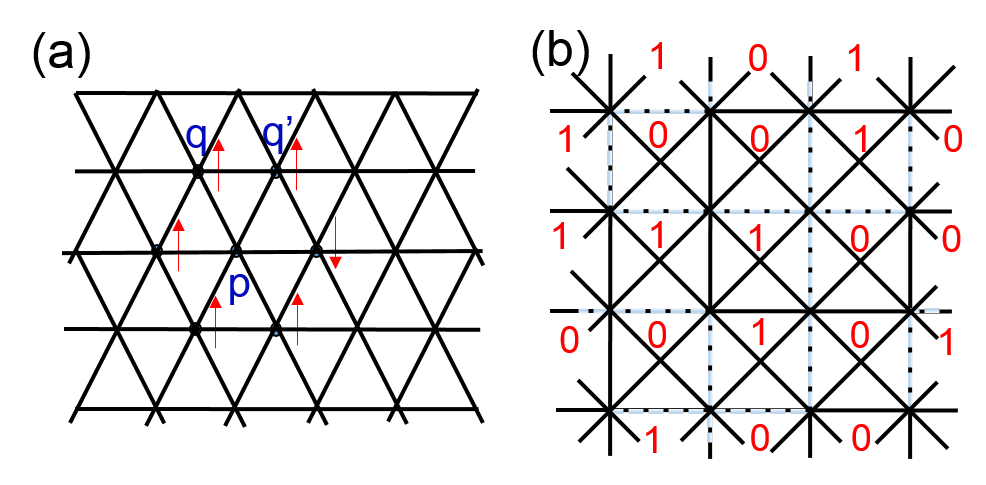}
  \caption{\label{fig:triangularUnionJack}
(a) Triangular lattice; (b) Union-Jack Lattice.}
\end{figure}

 Miller and Miyake have provided a fixed-point SPT qubit wave function that is invariant under $(Z_2)^3$ symmetry and is universal for MBQC~\cite{MillerMiyake16,MillerMiyake16b}. This wave function is defined on the Union-Jack lattice; see Fig.~\ref{fig:triangularUnionJack}. Due to the construction from 3-cocyles, this is manifestly a nontrivial SPT state. In particular, it is easily defined in terms of the so-called Control-Control-Z (CCZ) gate:
 \begin{equation}
 |\psi_{\rm MM}\rangle=\prod_{\langle p,q,r\rangle} {\rm CCZ}_{p,q,r} |++\cdots +\rangle,
 \end{equation}
 where the CCZ gates act on triplets of spins $p$, $q$, and $r$, sitting on vertices of triangles, and the action is
 \begin{eqnarray}
 {\rm CCZ}_{p,q,r}&\equiv& (|00\rangle\langle 00| +|01\rangle\langle 01|+|10\rangle\langle10|)_{pq}\otimes \openone_r \nonumber\\
 &&+|11\rangle\langle11|_{pq}\otimes Z_r.
 \end{eqnarray}
 As the CCZ gate is diagonal, it is symmetric under permutation of qubits $p$, $q$ and $r$. One can show that the Miller-Miyake state is ``stabilized'' by operators of the form $Q_p= \sigma_p^x \prod_{\langle pqr\rangle} CZ_{qr}$, where $\langle pqr\rangle$ denotes an triangle formed by vertices $p$, $q$ and $r$, i.e., $Q_p|\psi_{\rm MM}\rangle=|\psi_{\rm MM}\rangle$. Equivalently, one can regard a corresponding Hamiltonian such that $|\psi_{\rm MM}\rangle$ is the ground state,
 \begin{equation}
 H_{\rm MM}\equiv -\sum_{p}  \sigma_p^x \prod_{\langle pqr\rangle} CZ_{qr}.
 \end{equation}

\medskip\noindent {\bf Universal MBQC}. There are two approaches that Miller and Miyake used to the proved universality. One approach is to construct universal gates, and in this case, they were able to construct measurement patterns that yield both Toffoli and Hadamard gates, which together form a universal set of quantum gates. Another approach is to reduce the state, via local measurements, to some known universal resource state, which, in this case, is the graph state. We will explain the universality using this latter approach. 

To do this, observe that for the three qubits $p$, $q$ and $r$ that are acted by a CCZ gate, if one qubit, say $p$, is measured in the Z basis and the effective action on the remaining two qubits $q$ and $r$ is either identity $\openone_{qr}$ if the measurement outcome on $p$ is $|0\rangle$ or ${\rm CZ}_{qr}$ if the outcome is $|1\rangle$. Referring to Fig.~\ref{fig:triangularUnionJack}, we consider measurements on all the `cross' sites in the Z basis. An edge in the square sublattice has two such sites on opposite sides, and if the outcomes are different, i.e., one of them is $|0\rangle$ and the other is $|1\rangle$, then only one CZ gate acts on the two qubits (which were both in $|+\rangle$ state before the CCZ gates). This means that after Z measurements, the state of the remaining qubits form a graph state, whose graph depends on whether an edge is occupied (if one net CZ gate is applied), as indicated by the dashed edges in Fig.~\ref{fig:triangularUnionJack}b, or empty (if zero net CZ gate is applied). This becomes a bond percolation problem for the resultant graph state. If one can show that the resultant random graphs, with high probability, belong to the supercritical phase, then the original Miller-Miyake state on the Union-Jack lattice is a universal resource state. This is indeed what they showed in their percolation simulations~\cite{MillerMiyake16}. Miller and Miyake later generalized the construction to 3-cocycle states whose cocycle functions are tri-linear and each site can contains multiple qubits, and showed that these `fractional-symmetric' states are also universal~\cite{MillerMiyake16b}. 

The works by Miller and Miyake present exciting recent advancement in the connection of certain 2d SPT phases and MBQC. First, it can be realized by qubits, instead of higher-dimensional objects. Nevertheless, the parent Hamiltonian involves multiple-qubit interactions. Second, the requirement on the local measurement is greatly reduced: only Pauli measurements suffice.  Third, the quantum gates that can be achieved are in the 3rd-level of the Clifford hierarchy~\cite{Yoshida15}.   However, the universality may not exist on other simple lattices, such as the triangular lattice.  Again, a potential breakthrough is to establish an entire SPT phase supporting universal MBQC. 

\subsection{The Levin-Gu $Z_2$ model}
We introduced earlier the nontrivial 2d SPT order using the CZX model.  Levin and Gu constructed an interesting $Z_2$ symmetric Hamiltonian (invariant under the $Z_2$ symmetry $\prod_{p} \sigma_p^x$) on the triangular lattice [see Fig.~\ref{fig:triangularUnionJack}a],
\begin{eqnarray}
\label{eqn:LevinGu}
 &&H_{\rm LG}=-{\sum}_{p} \sigma_p^x S_p \equiv -{\sum}_p B_p,  \\
&& \mbox{where}\, S_p\equiv \prod_{\langle pqq'\rangle}S_{p;qq'}=\prod_{\langle pqq'\rangle} i^{({1-\sigma_q^z\sigma_{q'}^z})/{2}},
\end{eqnarray} where the product is over all triangles $\langle pqq'\rangle$  connected to site $p$, and despite its odd looking $S_p$ is a pure $\pm1$  factor depending on the spin configuration (in $\sigma^z$ basis) of the nearest neighboring 6 sites encircling the site $p$. This can be defined on other lattices with appropriate triangulation, such as the Union-Jack lattice. We now ask: do the the ground state of this Hamiltonian belong to the same phase of trivial paramagnet whose Hamiltonian is $H_{\rm trivial}=-\sum_p \sigma_p^x$\,? It was shown by Levin and Gu~\cite{LevinGu2012} that they represent two distinct 2D SPT phases (with $H_{\rm LG}$ giving a nontrivial phase), via gauging the global $Z_2$ symmetry and then examining the dual gauge models. It was also argued that since any gauge theory of a finite group $G$ is in one-to-one correspondence to an element in the third group cohomology $H^3[G,U(1)]$ the SPT phases  can also be understood via either this mathematical object. 

In fact one can write down the ground-state wave function of $H_{\rm LG}$ and it turns out it is closely related to the Miller-Miyake state that is defined on the same triangular lattice. They are also related to the cluster state, as we show below. 
\subsection{Connection of cluster state to Levin-Gu and Miller-Miyake models}

Here we reveal an intriguing relation between the Levin-Gu and Miller-Miyake models and to the cluster state. 
Let us define
\begin{equation}
U_{\rm S}\equiv\prod_{p}\prod_{\langle q,r\rangle\in\langle{p,q,r}\rangle } \big(|0\rangle_p\langle0|\otimes I_{qr} + |1\rangle_p\langle 1|\otimes S_{p; qr}\big).
\end{equation}
One can show that on a closed surface, this unitary transformation is the same as
 $U_{\rm CCZ}$, which is a product of all CCZ gates over all triangles $\langle p,q,r\rangle$ in the tiling,
 \begin{equation}
 \label{eqn:USUCCZ}
 U_{\rm S}=U_{\rm CCZ}=\prod_{\langle p,q,r\rangle} {\rm CCZ}_{p,q,r}.
 \end{equation}
 Using this, one can transform the $\sigma_x$ operator at site $p$ to
\begin{equation}
\label{eqn:UCCZ}
U_{\rm CCZ} \sum_p \sigma_x^{(p)} \,U_{\rm CCZ}^\dagger= \sum_p K_x^{(p)} S_p,
\end{equation} 
where $K_x^{(p)}=\sigma_x^{(p)} \otimes_{q\in {\rm Nb}(p)} \sigma_z^{(q)}$ is the usual stabilizer operator for the cluster state (defined on the triangular lattice in this case). Let us also define the product of all CZ gates as
$U_{\rm CZ}\equiv \prod_{\langle p,q\rangle\in E} {\rm CZ}_{pq}$,
where $\langle p,q\rangle$ belongs to an edge of the graph (in this case the triangular lattice). 

From the consideration of graph states and their stabilizer operators, we know that $U_{\rm CZ}$ can turn $K_x^{(p)}$ into $\sigma_x^{(p)}$, and hence,
\begin{equation}
\label{eqn:UCZUCCZ}
U_{\rm CZ} U_{\rm CCZ} \Big(-\sum_p \sigma_x^{(p)}\Big) \,U_{\rm CCZ}^\dagger U_{\rm CZ}^\dagger= -\sum_p \sigma_x^{(p)} S_p,
\end{equation} 
arriving at the Levin-Gu Hamiltonian. 
Now we are ready to state the relation between three different states: the cluster state $|\psi_{\rm cluster}\rangle$, the Levin-Gu state $|\psi_{\rm LG}\rangle$ (ground state of $H_{\rm LG}$) and the Miller-Miyake state $|\psi_{\rm MM}\rangle$ defined on the same lattice (tiled by triangles):
\begin{subequations}
\label{eqn:3eqs}
\begin{eqnarray}
&&|\psi_{\rm cluster}\rangle= U_{\rm CZ} |++\dots +\rangle, \\ 
&& |\psi_{\rm MM}\rangle=U_{\rm CCZ} |++\dots +\rangle,  \\
&&\ |\psi_{\rm LG}\rangle = U_{\rm CCZ} U_{\rm CZ} |++\dots +\rangle.
\end{eqnarray}
\end{subequations}

\smallskip \noindent {\bf Complement of edges in graphs}.
Due to the above relation, we have
\begin{equation}
|\psi_{\rm LG}\rangle= U_{\rm CZ} |\psi_{\rm MM}\rangle.
\end{equation}
Thus, we can infer that, the graph of the graph state after the measurement in the Miller-Miyake state is actually the `complement' of the graph after the same measurement outcome in the Levin-Gu state (due to additional CZ gates)! By `complement', we mean that the edges that were occupied become unoccupied, and vice versa. It would be interesting to see whether it can happen that the Miller-Miyake state on some particular lattice is not universal whereas the Levin-Gu state on the same lattice is, due to this complementary relation. It is also interesting to note that for all regular lattices the cluster state is universal, but with additional CCZ gates applied, the universality may disappear.

\section{Concluding remarks}

We have reviewed measurement-based approach for quantum computation, in which specific types of entanglement are known to provide a resource and only local measurement is needed. It remains one central open question as to what the complete characterization of universal resource states is. The study of computational universality in different families of states, such as the cluster/graph, the AKLT, and the SPT states,  and how it relates to physical properties may pave the road for such a cofmplete characterization.  Specifically,  whether an entire phase with certain 2D SPT order can be quantum computationally universal is an open question relevant to the notion of quantum computational phases of matter, for which the connection to the 1D SPT order has been substantiated recently~\cite{MillerMiyake,David1,David2}.  

Although cluster-state one-way computation is an abstract model of computation, it helps to make certain physical implementations more feasible. 
Knill, Laflamme and Milburn (KLM)~\cite{KLM} showed that quantum computation is possible with linear-optical elements with single-photon sources and detectors.  Nielsen later combined the KLM and the cluster-state approaches and developed an alternative optical quantum computation scheme that has less demanding resource requirement than KLM~\cite{NielsenCluster}. 
Browne and Rudolph developed another scheme without using teleported gates and achieved a more resource-efficient linear-optical quantum comptuation~\cite{BrowneRudolph}.  Cluster/graph states of small photons number have been nondeterministically generated (e.g. using downcoversion photons)~\cite{Walther,Lu}.
Another notable contribution of MBQC is the high error threshold in the surface-code quantum computation~\cite{RaussendorfHarrington,Fowler}.

There have also been schemes for deterministically generating cluster states of photons, e.g. using the coupling to quantum dots~\cite{LindnerRudolph,Economou}. Several important aspects of the schemes have recently been demonstrated~\cite{Schwartz}.
Certain graph states of phase-flip code were also created using trapped ions, with up to seven physical qubits~\cite{Lanyon}.

One of the most natural setting for MBQC is the cold-atom system~\cite{BlochReview}; creation of cluster-state cold atoms was realized some time ago~\cite{coldatoms}.
But the chanllenge is the local measurement; significant experimental progress has since been made that it is possible to detect and image single atoms~\cite{Greiner,Sherson,GreinerSimulation}.  One advantage of utilizing this system is the ability to scale up. Another 
 impressive achievement in realizing large-scale cluster states, from of order 100 to million modes~\cite{ChenMenicucciPfister,Yokoyama,Yoshikawa}, employs the continuous-variable quantum-optical system~\cite{Menicucci1,Menicucci2}.  The challenge to implement MBQC there is to perform  local optical-mode measurement.

Recent rapid development in quantum simulations using cold atoms, Rydberg atoms, trapped ions, cavities, photonics, and superconducting qubits, etc. have been proposed and developed to emulate spin Hamiltonians, and possible exotic Hamiltonians (such as topological orders) not necessarily existing naturally~\cite{ColdAtom,Exchange,OptLatt,Ryberg,Greiner,GreinerSimulation,Photonic,Cavities,Monroe,SC,Honeycomb}. If Hamiltonians of those universal resource ground states can be engineered, then the resource states can be created by `cooling' the system. An experimental simulation of such a cooling  was recently done for  cluster states~\cite{CoolCluster}. Alternatively, some resource states may be created by simple  acitve entangling procedure~\cite{coldatoms}.

Compared to cluster/graph states, other resource states are more of theoretical investigation and less experimental realization has been explored.
However, a short 1d AKLT chain was simulated in photons~\cite{Resch} and several quatum gates were demonstrated.
 For 2D, a proposal was made for the spin-3/2  state on the honeycomb lattice and for implementing the POVM~\cite{LiuLiGu}.  Small-scale entangled-photon  implementation is within the reach of current technology, but scaling up can be an issue. 
On the other hand it was recently proposed  to realize AKLT and general valence-bond states using $t_{2g}$ electrons in Mott insulators~\cite{Sela}. The ground state is  not the exact AKLT state, but in the same phase; thus the theoretical study of quantum computational universality away from the exact wavefunctions and even the entire phase is important.
SPT states in quantum computation is also mostly limited to theoretical consideration. However, it is possible that the above mentioned quantum simultations in various systems may provide useful avenue for realization.

\medskip \noindent {\bf Acknowledgment.}  The author acknowledges useful discussions with Ian Affleck, Akimasa Miyake, Abhishodh Prakash, Hendrik Poulsen Naturup Robert Raussendorf, David Stephen, Dongsheng Wang. This work was supported by the
National Science Foundation under Grant No. PHY
1620252.

\appendix
\section{Exercises}
\smallskip\noindent 
 1. Show that the CZ gate between two qubits $m$ and $n$ can be generated by the Hamiltonian $\hat{H}=(\openone_m- \sigma^z_m)(\openone_n- \sigma^z_n)$ for a duration $T=\pi/4$. 
 \smallskip \\ 2. Verify Eq.~(\ref{eqn:psi'}) and identify the omitted overall phase factor. How would the result change if we use $|\pm \xi\rangle=(e^{-i\xi/2}|0\rangle + e^{i\xi/2}|1\rangle)/\sqrt{2}$ instead?
\smallskip \\ 3. Verify Eq.~(\ref{eqn:U1}).
\smallskip \\ 4. Verify that ${\cal K}_k |{\cal C}_{\rm peri}\rangle=|{\cal C}_{\rm peri}\rangle$ and Eq.~(\ref{eqn:CZ-X}) using  ${\rm CZ}_{mn} X_m= X_m Z_n {\rm CZ}_{mn}$.
\smallskip \\ 5. Verify that Eq.~(\ref{eqn:Ku}) is a stablizer operator for the graph state $|G\rangle$.
\smallskip \\ 6. Fill in the missing steps and derive Eq.~(\ref{eqn:Gb}).
\smallskip  \\ 7. Verify the claim below Eq.~(\ref{eqn:xiPv}).
\smallskip  \\ 8. Verify Eq.~(\ref{eqn:PvAKLT}). 
\smallskip  \\ 9. Let $\vec{S}=\vec{S}_1 +\vec{S}_2$ for the two spin-1 particles, where $\vec{S}_i\cdot \vec{S}_i=1(1+1)$ with i=1,2. To construct a projector onto the join spin $S=2$ subspace, one can write 
  \begin{equation}
  P^{S=2}=a \big((\vec{S}_1+\vec{S}_2)^2-2\big)\big((\vec{S}_1+ \vec{S}_2)^2-0\big),
  \end{equation}
  which will annihilate any $S=0$ or $S=1$ state. Expand $P^{S=2}$ in terms of $\vec{S}_1\cdot \vec{S}_2$, show that the projector is proportional to the expression in the sum of Eq.~(\ref{eqn:HAKLT1D}). The constant $a$ is fixed by $P^{S=2}=a (\vec{S}^2-2)(\vec{S}^2-0)|_{\vec{S}^2=6}=I$.
 \smallskip   \\ 10. Verify the completeness relation Eq.~(\ref{eqn:FF}).
\smallskip \\   11. Verify the statement in the paragraph below Eq.~(\ref{eqn:1Dpost}) regarding reducing a three-site logical qubit to one single site. Namely,
     $a |\uparrow\downarrow\uparrow\rangle\otimes|\phi_0\rangle + b|\downarrow\uparrow\downarrow\rangle\otimes|\phi_1\rangle$ can be reduced to  $a |\uparrow\rangle\otimes|\phi_0\rangle + b|\downarrow\rangle\otimes|\phi_1\rangle$ (up to $\pm1$ relative phase dependent on measurement outcomes). 
 \smallskip   \\ 12. Further simplify Eq.~(\ref{eqn:UU}) and show that it gives a direct product of two single-qubit unitary gates, up to byproduct operators. 
 \smallskip \\  13. Further simplify Eq.~(\ref{eqn:CNOT}) and show that it gives the CNOT gate, up to byproduct operators. 
 \smallskip \\  14. Use a similar approach as in Exercise 9, to verify that the projector $P^{S=3}$ can be constructed using
      \begin{equation}
  P^{S=3}=b (\vec{S}\cdot \vec{S}-6)(\vec{S}\cdot \vec{S}-2)(\vec{S}\cdot \vec{S}-0),
  \end{equation}
and that it gives the Hamiltonian in Eq.~(\ref{eqn:HAKLT32}) up to an overall and an additive constants. Moreover, find out the overall constant $b$.
\smallskip \\ 15. Verify the completeness relation Eq.~(\ref{eqn:FF32}).
\smallskip \\ 16. Verify Eq.~(\ref{eqn:USUCCZ}),~(\ref{eqn:UCCZ}),~(\ref{eqn:UCZUCCZ}), and~(\ref{eqn:3eqs}).


\begin{thebibliography}{99}
\bibitem{NielsenChuang00}
M. Nielsen and I. Chuang, {Quantum Computation and Quantum Information\/}
(Cambridge Univ. Press, 2000).


\bibitem{Cluster}
H. J. Briegel and R. Raussendorf, 
Persistent Entanglement in Arrays of Interacting Particles, 
Phys. Rev. Lett. {\bf 86}, 910 (2001).

\bibitem{Oneway}
R. Raussendorf and H. J. Briegel, A One-Way Quantum Computer, Phys. Rev.
Lett. {\bf 86}, 5188-5191 (2001).


\bibitem{Oneway2}
H. J. Briegel, D. E. Browne, W. D\"ur, R. Raussendorf, and M. Van den Nest,
Measurement-based quantum computation, Nature Phys. {\bf 5}, 19-26 (2009).

\bibitem{ChildsLeungNielsen}
A. M. Childs, D. W. Leung, and M. A. Nielsen, {Unified
derivations of measurement-based schemes for quantum computation\/},
 Phy. Rev. A {\bf 71}, 032318 (2005).
 
\bibitem{GottesmanChuang}
D. Gottesman and I. L. Chuang,  {Demonstrating the viability of universal quantum computation using teleportation and single-qubit operations}, Nature {\bf 402}, 390 (1999).


\bibitem{Verstraete}
F. Verstraete and J. I. Cirac, Valence-bond states for quantum
computation, Phys. Rev. A {\bf 70} 060302(R) (2004).


\bibitem{Gross}
D. Gross and J. Eisert, Novel Schemes for Measurement-Based Quantum
Computation, Phys. Rev. Lett. {\bf 98}, 220503 (2007).


\bibitem{Gross2}
D. Gross, J. Eisert, N. Schuch, and D. Perez-Garcia, {Measurement-based quantum computation beyond the one-way model\/},
Phys. Rev. A {\bf 76}, 052315 (2007).

\bibitem{VandenNest1}
M. Van den Nest, W. D\"ur, G. Vidal, and H. J. Briegel, {Classical simulation versus universality in measurement-based quantum computation}, Phys. Rev.
A {\bf 75}, 012337 (2007).
\bibitem{VandenNest2}
M. Van den Nest, W. D\"ur, A. Miyake, and H. J. Briegel, {Fundamentals of universality in one-way quantum computation}, New J.
Phys. {\bf 9}, 204 (2007).

\bibitem{Gross1}
D. Gross, S.T. Flammia, and J. Eisert, ``Most Quantum States Are Too Entangled
To Be Useful As Computational Resources'', Phys. Rev. Lett. \textbf{102},
190501 (2009).

\bibitem{Bremner}
 M. J. Bremner, C. Mora, and A. Winter, {Are Random Pure States Useful for Quantum Computation?} Phys. Rev. Lett. {\bf 102}, 190502 (2009).
 
\bibitem{ChenGuWen2}
X. Chen, Z.-C. Gu, and X.-G. Wen, {Local unitary transformation,
long-range quantum entanglement, wave function renormalization, and
topological order\/}, Phys. Rev. B {\bf 82}, 155138 (2010).

 \bibitem{WenTO}
X.-G. Wen, {Topological Orders in Rigid States\/}, Int. J. Mod. Phys. B {\bf 4}, 239 (1990).

\bibitem{toric}
A. Y. Kitaev, {Fault-tolerant quantum computation by anyons\/},
Ann. of Phys. {\bf 303}, 2 (2003).



\bibitem{Balents}
L. Balents, {Spin liquids in frustrated magnets\/}, Nature {\bf 464}, 199 (2010).


\bibitem{TsuiStormerGossard}
D. C. Tsui, H. L. Stormer, and A. C. Gossard, {Two-Dimensional Magnetotransport in the Extreme Quantum Limit\/}, Phys. Rev. Lett. {\bf 48}, 1559 (1982).



\bibitem{Laughlin}
 R. B. Laughlin, {Anomalous Quantum Hall Effect: An Incompressible Quantum Fluid with Fractionally Charged Excitations\/}, Phys. Rev. Lett. {\bf 50}, 1395 (1983). 
 
 
\bibitem{TQC}
C. Nayak, S. H. Simon, A. Stern, M. Freedman, and S. Das Sarma, Non-Abelian anyons and topological quantum computation, Rev. Mod. Phys. {\bf 80}, 1083 (2008).  

\bibitem{BravyiRaussendorf}
S. Bravyi and R. Raussendorf, Measurement-based quantum computation with the toric code states,
Phys. Rev. A {\bf 76}, 022304 (2007).

\bibitem{RaussendorfAnn}
R. Raussendorf, J. Harrington, and K. Goyal, {A fault-tolerant one-way quantum computer\/}, Ann. Phys. {\bf 321}, 2242 (2006). 

\bibitem{RaussendorfNJP}
R. Raussendorf, J. Harrington, and K. Goyal, {Topological fault-tolerance in cluster state quantum computation\/},  New J. Phys. {\bf 9}, 199 (2007).

\bibitem{RaussendorfHarrington}
R. Raussendorf and J. Harrington, {Fault-tolerant quantum computation with high threshold in two dimensions\/}, Phys. Rev. Lett. {\bf 98}, 190504 (2007).



\bibitem{Fowler}

A. G. Fowler, M. Mariantoni, J. M. Martinis, and A. N. Cleland, Surface codes: Towards practical large-scale quantum computation,
Phys. Rev. A {\bf 86}, 032324 (2012).

\bibitem{Morimae}
T. Morimae, {Quantum computational tensor network on string-net condensate\/}, Phys. Rev. A {\bf 85}, 062328 (2012).

\bibitem{ElseSchwarzBartlettDoherty}
D. V. Else, I. Schwarz, S. D. Bartlett, and A. C. Doherty,
Symmetry-Protected Phases for Measurement-Based Quantum
Computation,
Phys. Rev. Lett. {\bf 108}, 240505 (2012).


\bibitem{MillerMiyake}
J. Miller and A. Miyake, Resource Quality of a Symmetry-Protected Topologically Ordered Phase for Quantum Computation, Phys. Rev. Lett. {\bf 114}, 120506 (2015).

 \bibitem{MillerMiyake16}
 J. Miller and A. Miyake, Hierarchy of universal entanglement in 2D measurement-based quantum computation, NPJ Quantum Information {\bf 2}, 16036 (2016).
 
\bibitem{Hendrik}
H. Poulsen Nautrup and T.-C. Wei, Symmetry-protected topologically ordered states for universal quantum computation, Phys. Rev. A {\bf 92}, 052309 (2015).

 \bibitem{MillerMiyake16b}
 J. Miller and A. Miyake, Latent Computational Complexity of Symmetry-Protected Topological Order with Fractional Symmetry, Phys. Rev. Lett. {\bf 120}, 170503 (2018).
 
\bibitem{WeiHuang}
T.-C. Wei and C.-Y. Huang, Universal measurement-based quantum computation in two-dimensional SPT phases, Phys. Rev. A {\bf 96}, 032317 (2017). 

\bibitem{RaussendorfWei}
R. Raussendorf and T.-C. Wei, {Quantum Computation by
Local Measurement}, Annu. Rev. Condens. Matter Phys. {\bf 3}, 239 (2012).

\bibitem{Kwek}
L. C. Kwek, Z. H. Wei, and B. Zeng. {Measurement-Based Quantum Computing with Valence-Bond-Solids}, Int. J. Mod. Phys. B {\bf 26}, 123002 (2012).

\bibitem{Fujii}
K. Fujii, Quantum Computation with Topological Codes: from qubit to topological fault-tolerance, arXiv:1504.01444 (2015).


\bibitem{Nielsen}
M. A. Nielsen,  Cluster-State Quantum Computation, Rep. Math. Phys. {\bf
57}, 147-161 (2005).

\bibitem{Hein}
M. Hein, W. Dur, J. Eisert, R. Raussendorf, M. Van den Nest, and H.-J. Briegel, in Quantum Computers, Algorithms and Chaos, International School of Physics Enrico Fermi Vol. 162 edited by G. Casati, D. Shepelyansky, P. Zoller, and G. Benenti (IOS Press, 2006).

\bibitem{Elliott}
M. B. Elliott, B. Eastin, C. M. Caves,
Graphical description of Pauli measurements on stabilizer states,
J. Phys. A: Math. Theor. {\bf 43}, 025301 (2010).

\bibitem{PEPS}
F. Verstraete, J.I. Cirac, V. Murg,
Matrix Product States, Projected Entangled Pair States, and variational renormalization group methods for quantum spin systems,
 Adv. Phys. {\bf 57}, 143 (2008).
 
\bibitem{MPS1}
S. Ostlund and S. Rommer, Thermodynamic Limit of Density Matrix
Renormalization, Phys. Rev. Lett. {\bf 75}, 3537 (1995).
\bibitem{MPS2}
M. Fannes, B. Nachtergaele, and R. F. Werner, Finitely correlated states on quantum spin chains, Commun. Math.
Phys. {\bf 144}, 443 (1992).


\bibitem{AKLT}
I. Affleck, T. Kennedy, E. H. Lieb, and H. Tasaki, {Rigorous
results on valence-bond ground states in antiferromagnets\/}, Phys.
Rev. Lett. {\bf 59}, 799 (1987).

\bibitem{AKLT2}
I. Affleck, T. Kennedy, E. H. Lieb, and H. Tasaki, {Valence bond
ground states in isotropic quantum antiferromagnets\/},  Comm. Math.
Phys. {\bf 115}, 477 (1998).

\bibitem{Haldane}
F. D. M. Haldane, {Continuum dynamics of the 1-D Heisenberg
antiferromagnet: Identification with the O(3) nonlinear sigma
model\/}, Phys. Lett. A {\bf 93}, 464 (1983).

\bibitem{Haldane2}
F. D. M. Haldane, {Nonlinear Field Theory of Large-Spin
Heisenberg Antiferromagnets: Semiclassically Quantized Solitons of
the One-Dimensional Easy-Axis N\'eel State\/}, Phys. Rev. Lett. {\bf
50}, 1153 (1983).


 \bibitem{Brennen}
 G. K. Brennen and A. Miyake, Measurement-Based Quantum Computer in the Gapped Ground State of a Two-Body Hamiltonian, Phys. Rev. Lett. {\bf 101}, 010502
 (2008).
 
 \bibitem{ChenZeng}

X. Chen, R. Duan, Z. Ji, and B. Zeng, Quantum State Reduction for Universal Measurement Based Computation,
Phys. Rev. Lett. {\bf 105}, 020502 (2010).

\bibitem{Knabe}
S. Knabe, {Energy gaps and elementary excitations for certain
VBS-quantum antiferromagnets\/}, J. Stat. Phys. {\bf 52}, 627
(1988).

  \bibitem{WeiAffleckRaussendorf11}
 T.-C. Wei, I. Affleck, and R. Raussendorf,
 Affleck-Kennedy-Lieb-Tasaki State on a Honeycomb Lattice is a Universal Quantum Computational Resource,
  Phys. Rev. Lett. {\bf 106}, 070501 (2011).
  
\bibitem{WeiAffleckRaussendorf12}
 T.-C. Wei, I. Affleck, and R. Raussendorf, Two-dimensional Affleck-Kennedy-Lieb-Tasaki state on the honeycomb lattice is a universal resource for quantum computation,
 Phys. Rev. A {\bf 86}, 032328 (2012).
 
\bibitem{SPTRG}
Z.-C. Gu, and X.-G. Wen, Tensor-entanglement-filtering renormalization approach and symmetry protected topological order Phys. Rev. B \textbf{80} 155131 (2009).

\bibitem{1DSPTcomplete}
X. Chen, Z.-C. Gu, and X.-G. Wen, Complete classification of 1D gapped quantum phases in interacting spin systems, Phys. Rev. B \textbf{84}, 235128 (2011).

\bibitem{PollmannSPT}
 F. Pollmann, E. Berg, A. M. Turner, and M. Oshikawa, Symmetry protection of topological order in one-dimensional quantum spin systems , Phys. Rev. B {\bf 85}, 075125 (2012).
 
\bibitem{SPTtensor}
N. Schuch, D. P\'erez-Garc\'ia, and J. I. Cirac, Classifying quantum phases using MPS and PEPS, Phys. Rev. B \textbf{84}, 165139 (2011).

\bibitem{ChenScience}
X. Chen, Z.-C. Gu, Z.-X. Liu, and X.-G. Wen, Symmetry-Protected Topological Orders in Interacting Bosonic Systems, Science {\bf 338},
1604 (2012).


\bibitem{cohomology} X. Chen, Z.-C. Gu, Z.-X. Liu, and X.-G. Wen, Symmetry protected topological orders and the group cohomology of their symmetry group, Phys. Rev. B {\bf 87}, 155114 (2013).

\bibitem{Abhi}
A. Prakash and T.-C. Wei, Ground-state forms of 1D symmetry-protected topological phases and their utility as resource states for measurement-based quantum computation, Phys. Rev. A {\bf 92}, 022310 (2015).

\bibitem{David1}
David T. Stephen, Dong-Sheng Wang, Abhishodh Prakash, Tzu-Chieh Wei, and Robert Raussendorf,  Computational power of symmetry protected topological phases, Phys. Rev. Lett. {\bf 119}, 010504 (2017).
 
 \bibitem{David2}
 Robert Raussendorf, Dongsheng Wang, Abhishodh Prakash, Tzu-Chieh Wei, and David Stephen, Symmetry-protected topological phases with uniform computational power in one dimension, Phys. Rev. A {\bf 96}, 012302 (2017).
 
\bibitem{BriegelRaussendorf}
H. J. Briegel and R. Raussendorf, Persistent Entanglement in arrays of Interacting Particles, Phys. Rev. Lett. {\bf 86}, 910 (2001).

\bibitem{BlindQC}

A. Broadbent, J. Fitzsimons, and E. Kashefi, Universal blind quantum computation,
Proceedings of the 50th Annual IEEE Symposium on Foundations of Computer Science (FOCS 2009), pp. 517-526

\bibitem{BlindQCScience}

S. Barz, E. Kashefi, A. Broadbent, J. F. Fitzsimons, A. Zeilinger, and P. Walther, Demonstration of Blind Quantum Computing,
Science  {\bf 335},  pp. 303-308
(2012).

\bibitem{Browne}
D. E. Browne, M. B. Elliott, S. T. Flammia, S. T. Merkel, A. Miyake, and A. J. Short, Phase transition of computational power in the resource states for one-way quantum computation, New J. Phys. {\bf 10}, 023010 (2008).

\bibitem{Haselgrove}

H. L. Haselgrove, M. A. Nielsen, and T. J. Osborne, Quantum States far from the Energy Eigenstates of Any Local Hamiltonian,
Phys. Rev. Lett. {\bf 91}, 210401 (2003).
\bibitem{ChenChenDuanJiZeng}
J. Chen, X. Chen, R. Duan, Z. Ji, and B. Zeng, No-go theorem for one-way quantum computing on naturally occurring two-level systems,
Phys. Rev. A {\bf 83}, 050301(R) (2011).
\bibitem{Bravyi}
S. Bravyi, Efficient algorithm for a quantum analogue of 2-SAT, arXiv:quant-ph/0602108.
\bibitem{JiWeiZeng}
Z. Ji, Z. Wei, and B. Zeng, Complete characterization of the ground-space structure of two-body frustration-free Hamiltonians for qubits,
Phys. Rev. A {\bf 84}, 042338 (2011).

\bibitem{RudolphBartlett}
S. D. Bartlett and T. Rudolph, Simple nearest-neighbor two-body Hamiltonian system for which the ground state is a universal resource for quantum computation, Phys. Rev. A {\bf 74}, 040302(R) (2006).

 \bibitem{Tricluster}
 X. Chen, B. Zeng, Z.-C. Gu, B. Yoshida, and I. L. Chuang, Gapped Two-Body Hamiltonian Whose Unique Ground State Is Universal
for One-Way Quantum Computation, Phys. Rev. Lett. {\bf 102}, 220501 (2009).
\bibitem{VBS}
N. Read and S. Sachdev, Valence-bond and spin-Peierls ground states of low-dimensional quantum antiferromagnets, Phys. Rev. Lett. {\bf 62}, 1694 (1989).
\bibitem{ParameswaranSondhiArovas09}
 S. A. Parameswaran, S. L. Sondhi, and D. P. Arovas,
{Order and disorder in AKLT antiferromagnets in three
dimensions\/}, Phys. Rev. B {\bf 79}, 024408 (2009).

\bibitem{spinglass}
C. R. Laumann, S. A. Parameswaran, S. L. Sondhi, and F. Zamponi,
{AKLT Models with Quantum Spin Glass Ground States\/}, Phys.
Rev. B {\bf 81}, 174204 (2010).

 \bibitem{Miyake11} A. Miyake, Quantum computational capability of a two-dimensional valence bond solid phase,
Ann. Phys. (Leipzig) {\bf 326}, 1656 (2011).

\bibitem{Wei13}

T.-C. Wei, Quantum computational universality of spin-3/2 Affleck-Kennedy-Lieb-Tasaki states beyond the honeycomb lattice, Phys. Rev. A {\bf 88}, 062307 (2013).

\bibitem{WeiPoyaRaussendorf}
T.-C. Wei, P. Haghnegahdar, and R. Raussendorf, Hybrid valence-bond states for universal quantum computation, Phys. Rev. A {\bf 90}, 042333 (2014). 


\bibitem{WeiRaussendorf15}
T.-C. Wei and R. Raussendorf, Universal measurement-based quantum computation with spin-2 Affleck-Kennedy-Lieb-Tasaki states, Phys. Rev A {\bf 92}, 012310 (2015).

\bibitem{HuangWagnerWei}
C.-Y. Huang, M. A. Wagner, and T.-C. Wei, Emergence of the XY-like phase in the deformed spin-3/2 AKLT systems, Phys. Rev. B {\bf 94}, 165130 (2016).

\bibitem{AKLTgap}
A. Garcia-Saez, V. Murg, T.-C. Wei, {Spectral gaps of Affleck-Kennedy-Lieb-Tasaki Hamiltonians using Tensor Network methods\/},
Phys. Rev. B {\bf 88}, 245118 (2013).

\bibitem{VerstraeteGroup}

L. Vanderstraeten, M. Mari\"n, F. Verstraete, and J. Haegeman, Excitations and the tangent space of projected entangled-pair states,
Phys. Rev. B {\bf 92}, 201111(R) (2015).

\bibitem{NiggemannKlumperZittarz}
H. Niggemann, A. Kl\"umper, and J. Zittartz, 
{Quantum phase transition in spin-3/2 systems on the hexagonal lattice--- optimum
ground state approach}, 
Z. Phys. B {\bf 104}, 103 (1997).

\bibitem{NiggemannZittartz}
H. Niggemann and J. Zittartz, Ground state properties of a spin-3/2 model on a decorated square lattice, Eur. Phys. J. B {\bf 13}, 377 (2000).

\bibitem{NKZspin2}
H. Niggemann, A. Kl\"umper, and J. Zittartz, Ground state phase diagram of a spin-2 antiferromagnet on the square lattice, Eur.
Phys. J. B {\bf 13}, 15 (2000).
  \bibitem{DarmawanBrennenBartlett}
 A. S. Darmawan, G. K. Brennen, and S. D. Bartlett,
 Measurement-based quantum computation in a 2D phase of matter,
 New J. Physics, {\bf 14}, 013023 (2012).

\bibitem{Darmawan2}
A. S. Darmawan and S. D. Bartlett, Graph states as ground states of two-body frustration-free Hamiltonians, New J. Phys. {\bf  16}, 073013 (2014).

\bibitem{DarmawanBartlett16}

A. S. Darmawan and S. D. Bartlett, Spectral properties for a family of two-dimensional quantum antiferromagnets,
 Phys. Rev. B {\bf 93}, 045129 (2016).
\bibitem{CZX} X. Chen, Z.-X. Liu, and X.-G. Wen, Two-dimensional symmetry-protected topological orders and their protected gapless edge excitations, Phys. Rev. B {\bf 84}, 235141 (2011).


\bibitem{Yoshida15}
B. Yoshida, Gapped boundaries, group cohomology and fault-tolerant logical gates, Annals of Physics {\bf 337}, 387 (2017).

\bibitem{LevinGu2012}
M. Levin and Z.-C. Gu, {Braiding statistics approach to symmetry-protected topological phases}, Phys. Rev. B {\bf 86}, 115109 (2012).


\bibitem{KLM}
E. Knill E, R. Laflamme, and G. J. Milburn, A scheme for efficient quantum computation with linear optics, Nature {\bf 409}, 46–52 (2001).

\bibitem{NielsenCluster}
M. A. Nielsen, Optical Quantum Computation Using Cluster States, Phys. Rev. Lett. {\bf 93}, 040503 (2004).

\bibitem{BrowneRudolph}
D. E. Browne and T. Rudolph, Resource-Efficient Linear Optical Quantum Computation, Phys. Rev. Lett. {\bf 95}, 010501 (2005).

\bibitem{Walther}
P. Walther, K. J. Resch, T. Rudolph, E. Schenck, H. Weinfurter, V.
Vedral, M. Aspelmeyer and A. Zeilinger, 
Experimental one-way quantum computing, 
Nature {\bf 434}, 169 (2005).



\bibitem{Lu}
C.-Y. Lu et al., Experimental entanglement of six photons in graph states, Nature Physics {\bf 3}, 91 - 95 (2007).




  \bibitem{LindnerRudolph}  
N. H. Lindner, T. Rudolph, Proposal for Pulsed On-Demand Sources of Photonic Cluster State Strings, Phys. Rev. Lett. {\bf 103}, 113602
(2009).

\bibitem{Economou}
S. E. Economou, N. Lindner, T. Rudolph, Optically Generated 2-Dimensional Photonic Cluster State from Coupled Quantum Dots, Phys. Rev. Lett. {\bf 105},
093601 (2010).



\bibitem{Schwartz}
I. Schwartz et al., Deterministic generation of a cluster state of entangled photons, Science {\bf 354}, 434 (2016).






\bibitem{Lanyon}
B. P. Lanyon et al., Measurement-Based Quantum Computation with Trapped Ions, Phys. Rev. Lett. {\bf 111}, 210501 (2013)







\bibitem{BlochReview}
I. Bloch, Quantum coherence and entanglement with ultracold atoms in optical lattices, Nature {\bf 453}, 1016 (2008).

\bibitem{coldatoms}
 O. Mandel, M. Greiner, A. Widera, T. Rom, T. W.
H\"ansch, and I. Bloch, Controlled collisions for multi-particle entanglement of optically trapped atoms,
Nature
{\bf 425}, 937 (2003).

\bibitem{Greiner}
W. S. Bakr, J. I. Gillen, A. Peng, S. F\"olling, and M. Greiner,
A quantum gas microscope for detecting single atoms in a Hubbard-regime optical lattice,
Nature {\bf 462}, 74 (2009).
%

\bibitem{Sherson}
J. F. Sherson, 	C. Weitenberg,	M. Endres,	M. Cheneau,	I. Bloch, and S. Kuhr, 
Single-atom-resolved fluorescence imaging of an atomic Mott insulator. 
Nature {\bf 467}, {68} (2010).


\bibitem{GreinerSimulation}
J. Simon,	W. S. Bakr,	R. Ma,	M. E. Tai,	P. M. Preiss, and M. Greiner, 
{Quantum simulation of antiferromagnetic spin chains in an optical lattice\/},
Nature {\bf 472}, 307 (2011).

\bibitem{ChenMenicucciPfister}
 M. Chen, N. C. Menicucci, and O. Pfister,
 Experimental realization of multipartite entanglement of 60 modes of a quantum optical frequency comb, 
 Phys. Rev. Lett. {\bf 112}, 120505 (2014).
 
\bibitem{Yokoyama}
 S. Yokoyama, R. Ukai, S. C. Armstrong, C. Sornphiphatphong, T. Kaji, S. Suzuki, J. Yoshikawa, H. Yonezawa, N. C.
Menicucci, and A. Furusawa, 
Ultra-large-scale continuous-variable cluster states multiplexed in the time domain,
Nat. Photonics {\bf 7}, 982 (2013).
\bibitem{Yoshikawa}
J.-i. Yoshikawa, S. Yokoyama, T. Kaji, C. Sornphiphatphong, Y. Shiozawa, K. Makino, and A. Furusawa, Generation of one-million-mode continuous-variable cluster state by unlimited time-domain multiplexing dual-rail continuous variable cluster state, APL Photonics {\bf 1}, 060801 (2016).



\bibitem{Menicucci1}
N. C. Menicucci, S. T. Flammia, H. Zaidi, and O. Pfister, 
Ultracompact generation of continuous-variable cluster states,
Phys. Rev. A {\bf 76}, 010302(R) (2007).

\bibitem{Menicucci2}
 N. C. Menicucci, S. T. Flammia, and O. Pfister, 
One-way quantum computing in the optical frequency comb,
Phys. Rev.
Lett. {\bf 101}, 130501 (2008).

\bibitem{ColdAtom}
L.-M. Duan, E. Demler, and M. D. Lukin, 
{Controlling Spin Exchange Interactions of Ultracold Atoms in Optical Lattices\/},
Phys. Rev. Lett. {\bf 91}, 090402 (2003).

\bibitem{Exchange}

M. Anderlini, P. J. Lee, B. L. Brown, J. Sebby-Strabley, W. D. Phillips, and J. V. Porto,
{Controlled Exchange Interaction Between Pairs of Neutral Atoms in an Optical Lattice\/},
Nature {\bf 448},  452-456 (2007).

\bibitem{OptLatt}
M. Aguado, G. K. Brennen, F. Verstraete, J. I. Cirac,
{Creation, manipulation, and detection of Abelian and non-Abelian anyons in optical lattices\/}, 
Phys. Rev. Lett. {\bf 101}, 260501 (2008).

\bibitem{Ryberg}
H. Weimer, M. M\"uller, I. Lesanovsky, P. Zoller, and  H. P. B\"uchler, 
{A Rydberg quantum simulator\/}, 
Nature Phys.  {\bf 6}, 382-388 (2010).

\bibitem{Photonic}
X.-S. Ma, B. Dakic, S. Kropatschek, W. Naylor, Y.-H. Chan, Z.-X. Gong, L.-M. Duan, A. Zeilinger, and P. Walther, 
{Towards photonic quantum simulation of ground states of frustrated Heisenberg spin systems\/},
Scientific Reports {\bf 4},  3583 (2014).

\bibitem{Cavities}
X.-W. Luo,	X. Zhou,	C.-F. Li,	J.-S Xu,	G.-C. Guo, and Z.-W. Zhou, 
{Quantum simulation of 2D topological physics in a 1D array of optical cavities\/},
Nature Communications {\bf 6},  7704 (2015). 

\bibitem{Monroe}
C. Senko, P. Richerme, J. Smith, A. Lee, I. Cohen, A. Retzker, and C. Monroe,
{Realization of a Quantum Integer-Spin Chain with Controllable Interactions\/}, 
Phys. Rev. X {\bf 5}, 021026 (2015).

\bibitem{SC}


A. A. Houck,	H. E. Türec, and  J. Koch, 
{On-chip quantum simulation with superconducting circuits\/},
Nature Phys. {\bf 8}, 292 (2012).

\bibitem{Honeycomb}
G. Jotzu,	M. Messer,	R. Desbuquois,	M. Lebrat, 	T. Uehlinger, D. Greif, and T. Esslinger, 
{Experimental realization of the topological Haldane model with ultracold fermions\/}, 
Nature {\bf 515}, 237 (2014).

\bibitem{CoolCluster}
G. H. Aguilar, T. Kolb, D. Cavalcanti, L. Aolita, R. Chaves, S. P. Walborn, and P. H. Souto Ribeiro,
Linear-Optical Simulation of the Cooling of a Cluster-State Hamiltonian System,
Phys. Rev. Lett. {\bf 112}, 160501 (2014).


\bibitem{Resch}
R. Kaltenbaek,   J. Lavoie,    B. Zeng,   S. D. Bartlett, and K. J.
Resch, 
{Optical one-way quantum computing with a simulated valence-bond solid\/}, 
Nature Phys. {\bf 6}, 85 (2010).

\bibitem{LiuLiGu}
K. Liu, W.-D. Li, and Y.-J. Gu, 
Measurement-based quantum computation with an optical two-dimensional Affleck–Kennedy–Lieb–Tasaki state,
J. Opt. Soc. Am. B {\bf 31}, 2689 (2014).

\bibitem{Sela}
M. Koch-Janusz, D. I. Khomskii, E. Sela, 
Two-dimensional Valence Bond Solid (AKLT) states from $t_{2g}$ electrons,
Phys. Rev. Lett. {\bf 114}, 247204 (2015).





\end{thebibliography}
\end{document}